\documentclass[11pt]{article}
\usepackage[left=0.75in,right=0.75in,top=0.75in,bottom=1.0in]{geometry}
\usepackage{amsmath,amsfonts,amssymb,bm}
\usepackage{mathtools}
\usepackage{graphicx}
\usepackage{booktabs}
\usepackage{float}
\usepackage{subcaption} 
\newcommand{\vect}[1]{\bm{#1}}

\usepackage{placeins}
\usepackage[numbers,sort&compress]{natbib}
\usepackage{etoolbox}

\AtBeginEnvironment{thebibliography}{%
  \small                             
  \setlength{\parskip}{0pt}%
  \setlength{\itemsep}{2pt plus 0.3ex}%
  \setlength{\parsep}{0pt}%
}
\usepackage[hidelinks]{hyperref}

\usepackage{authblk}
\title{Harnessing Floating Car Data, Traffic Camera Observations, and Network Flow Analysis for Traffic Volume Estimation}
\author[1]{Antonina Kosikova\thanks{This manuscript is a preprint version of a work submitted to Discover Civil Engineering.}}
\author[1]{Mehmet Kerem Turkcan}
\author[2]{Ahmed Darrat}
\author[1]{Andrew Smyth}
\affil[1]{Department of Civil Engineering and Engineering Mechanics, Columbia University, New York, NY 10027, USA}
\affil[2]{INRIX, Bellevue,  Washington, 9800, USA}

\date{} 
\begin{document}
\maketitle
\newcommand{\keywords}[1]{%
  \vspace{0.75\baselineskip}%
  \noindent\textbf{Keywords:} #1%
}
\begin{abstract}
Cities increasingly rely on vehicle trajectory data to monitor traffic conditions; however, such data offer only a partial and spatially heterogeneous view of network dynamics and exhibit systematic biases across corridors and time periods.  In contrast, surveillance cameras can provide high-fidelity traffic information, but only at a limited set of locations, typically sparsely distributed across the road network. We present a hybrid modeling and calibration framework that fuses these complementary data sources to produce physically consistent, network-wide estimates and short-horizon forecasts of traffic volumes. The framework leverages kinematic features derived from the Cell Transmission Model (CTM) formulation within a graph neural network (GNN). By enforcing traffic-flow conservation, capacity limits, and spillback dynamics, the CTM provides a physically grounded representation of traffic flow, while the GNN learns the spatiotemporal evolution of traffic states over the entire road network.
To calibrate the model predictions on traffic camera observations, we use a progressive data-assimilation scheme based on an Ensemble Square-Root Kalman filter (EnSRF). A topology-informed flow-weighted transition matrix is further employed to propagate camera-driven corrections to unobserved road segments, enabling real-time, network-wide traffic state and volume estimation. The approach is demonstrated using probe-vehicle trajectory data and municipal traffic cameras in Manhattan, New York City, where it achieves improved accuracy relative to trajectory-based estimates while maintaining physically plausible and network-consistent traffic flows.
The proposed framework accommodates varying sensor availability and produces calibrated traffic volumes with uncertainty estimates, supporting operational monitoring and evaluation of transportation policies in data-constrained urban environments.
\end{abstract}

\keywords{Floating Car Data, Traffic State Estimation, Graph Neural Network, Ensemble Square-Root Kalman filter}

\maketitle

\section{Introduction}\label{Sec.1}
Reliable urban traffic prediction is essential for transportation operations, emergency response, and long-term infrastructure planning. In recent years, transportation agencies have increasingly relied on vehicle trajectory data, also known as floating-car data,  derived from GPS-equipped probe vehicles, mobile devices, or fleet services, to infer network-wide traffic conditions, owing to their broad spatial coverage and fine temporal resolution. However, the probe-vehicle data are inherently undersampled with penetration rates that vary significantly across locations, time periods, and travel modes \cite{herrera2010evaluation, freeway}. This spatially and temporally heterogeneous coverage leads to biased and incomplete representations of traffic states, particularly in urban environments with a prevalence of low traffic volume arterial roads \cite{work2008ensemble}.  To address this challenge, traffic state estimation methods have incorporated fixed sensors, such as loop detectors \cite{loop}, license-plate recognition systems \cite{license}, and radar sensors \cite{czyzewski2019estimating}. Advances in communication technologies within transportation systems have further expanded these capabilities by enabling the use of surveillance video systems for traffic data collection \cite{xue2012real, ide2016city, wang2018real}. Although such developments have opened new opportunities for accurate traffic monitoring, the high installation and maintenance costs associated with traffic cameras continue to restrict their widespread deployment in dense urban environments populated with arterial roads.  As a result, traffic camera coverage is typically concentrated on highways and at toll collection points, which in turn motivated studies focused on traffic state estimation for freeway and highway systems \cite{seo2017traffic, wang2005real, work2008ensemble, freeway}. \\
Highway networks are typically characterized by relatively homogeneous traffic dynamics, constrained lane-changing behavior, and well-defined boundary conditions. Such settings permit the use of simplified macroscopic or mesoscopic models with fundamental diagrams and comparatively stable demand patterns \cite{treiber2010three}. While such assumptions are appropriate for controlled-access highways, they become increasingly restrictive in dense urban environments, where frequent intersections, signal control,  pedestrian and curbside operations, and irregular local demand introduce strong spatial heterogeneity and nonstationarity into traffic dynamics \cite{work2008ensemble}. To address traffic state estimation in urban networks, studies have commonly relied on simulation-based tools and probe-vehicle data, such as taxi fleets \cite{tang2019joint,liu2024spatial, herring2010estimating, wang2022inferring}, or commercial vehicle trajectories \cite{kim2015spatial}, complemented by fixed detectors and traffic camera information, which are often used interchangeably for both model development and validation \cite{kaiser2025spatio}. Although multiple data sources can provide valuable spatiotemporal information, current practices remain constrained by sparse and biased sensing, ad‑hoc treatment of heterogeneous data, and limited transferability beyond single network case studies.
In the absence of a unified framework for reliable traffic state estimation in urban environments, network-wide traffic volume estimation remains a challenging problem, which limits the practical deployment of existing approaches in real-world, city-scale settings, particularly when the probe data are sparse, biased toward specific vehicle classes, or insufficient to serve as an independent basis for model development. \\
In this work, we address the problem of urban traffic volume estimation by presenting a data-fusion framework designed for conditions of limited probe-vehicle data availability and sparse traffic camera coverage. To estimate latent 
traffic states and predict network-wide traffic volumes, we adopt a hierarchical strategy that integrates physics-based modeling, graph learning, and sequential data assimilation.
In the first stage, a Graph Neural Network (GNN) is trained on a Cell Transmission Model (CTM)-derived kinematic features extracted from low-penetration probe-vehicle trajectories. These features enable the model to learn spatiotemporal dependencies and baseline traffic correlations from the broad, albeit incomplete, coverage of the trajectory data. In the second stage, the GNN predictions are used as a dynamical prior for estimating time- and location-dependent calibration factors that relate the trajectory-derived volumes to the traffic camera observation using an Ensemble Square-Root Kalman Filter (EnSRF) \cite{tippett2003ensemble, Evensen2003_EnKF}. Temporal variation in these factors is modeled using covariates such as month, day of week, and hour of day. The resulting calibration information is then propagated to camera-unobserved network segments through a flow-weighted transition matrix that captures directional traffic coupling across the road network. By anchoring GNN learning to conservation-consistent CTM dynamics and calibrating biased trajectory measurements on camera observations, the framework produces physically consistent, network-wide traffic volume estimates and short-horizon forecasts under partial observability.
The performance of the proposed framework is demonstrated on the Manhattan road network in New York City under conditions of sparse traffic camera coverage,  where a subset of the available cameras is used to calibrate the trajectory-derived volumes, while the remaining cameras are reserved for independent model validation.

\section{Related Work}
GNNs have been progressively adopted in transportation research due to their ability to naturally represent the spatial connectivity and temporal dependencies inherent in traffic dynamics \cite{jin2023spatio}. Various studies have demonstrated the potential of GNNs for traffic state estimation and forecasting on both highway systems  \cite{bai2020adaptive,cai2020traffic}, and urban road networks \cite{ kaiser2025spatio,cui2019traffic}. Leveraging the capabilities of GNNs, a substantial body of prior work has also addressed the problem of recovering vehicle movements and estimating traffic states from incomplete observations \cite{zheng2012reducing, banerjee2014inferring}.  Consequently, GNNs have been used to integrate multiple sources of information, improving the representativeness of inferred network conditions.  In this context, the traffic cameras are considered the most accurate sensing modality for traffic volume estimation.  With the use of object-detection methods, camera streams can provide continuous measurements and distinguish vehicle classes \cite{sapkota2025yolo26}. 
Nevertheless, the utility of camera data for network-wide estimation is fundamentally constrained by their sparse spatial coverage, particularly in dense urban environments where sensor deployment is limited. 
However, most existing studies either assume relatively dense camera coverage over the road network or focus on highway corridors, where traffic conditions can be directly observed over a large fraction of the road network \cite{xue2012real, ide2016city, wang2018real}. 
Such settings are less representative of many U.S. cities, where camera deployment is often limited by privacy concerns, institutional constraints, and the high costs of maintenance. 
To address limited direct sensing, several studies have combined probe-vehicle trajectories with simulation or data-driven models. One such approach leverages taxi trajectory data and validates the model performance using the SUMO traffic simulator \cite{tang2019joint}. Another study uses SUMO to simulate the traffic dynamics and incorporates the surveillance camera records for citywide traffic volume inference \cite{yu2019citywide}. More recently, versatile frameworks, including variants of large language models (LLMs), have also been explored for traffic forecasting using datasets derived from taxis, bike-sharing systems, and related urban mobility platforms \cite{liu2024spatial}. \\
Importantly, despite extensive work on incorporating heterogeneous data sources into traffic state estimation, only a relatively small number of studies explicitly address uncertainty quantification in the inferred traffic states \cite{deng2013traffic}. The existing models typically employ Bayesian filtering, such as ensemble Kalman filtering and nonlinear Kalman variants such as the ensemble (EnKF) \cite{work2008ensemble} or unscented Kalman filter (UKF) \cite{ngoduy2011low}, that are used to provide uncertainty-aware traffic state estimates. More recently, deep probabilistic and generative models have also been used to learn predictive distributions over traffic states, leveraging likelihood-based loss functions and heteroscedastic regression approaches \cite{li2024real,ding2025uncertainty}.

\section{Problem statement}
\begin{figure}[t]
    \centering
    \begin{minipage}[t]{0.48\textwidth}
        \centering
        \vspace{0pt}
        \textbf{(a)}\\[0.3em]
        \includegraphics[width=\linewidth]{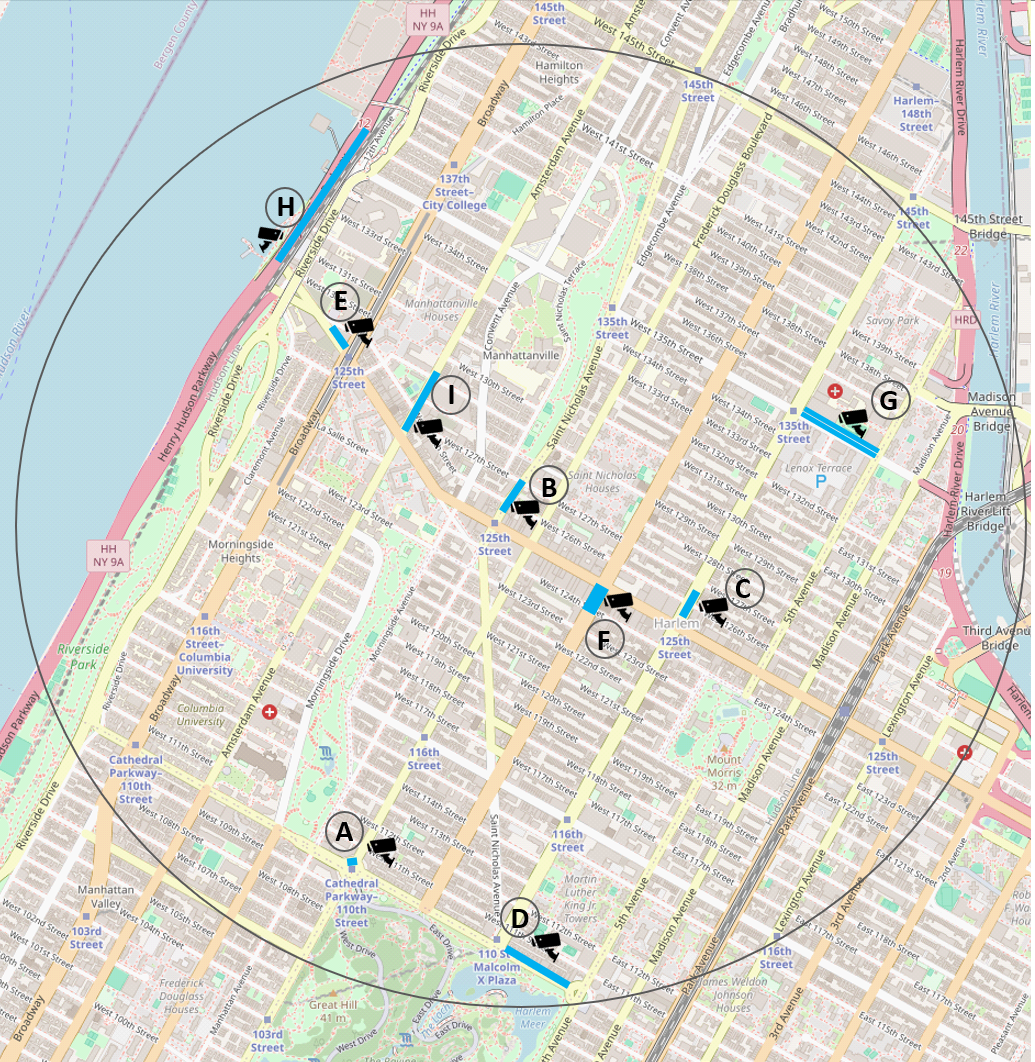}
    \end{minipage}
    \hspace{2em}
    \begin{minipage}[t]{0.35\textwidth}
        \centering
        \vspace{0pt}
        \textbf{(b)}\\[0.3em]
        \includegraphics[width=\linewidth]{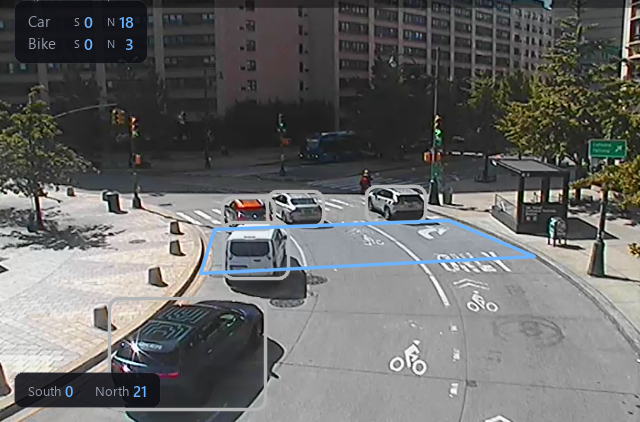}

        \vspace{1em}

        \textbf{(c)}\\[0.3em]
        \includegraphics[width=\linewidth]{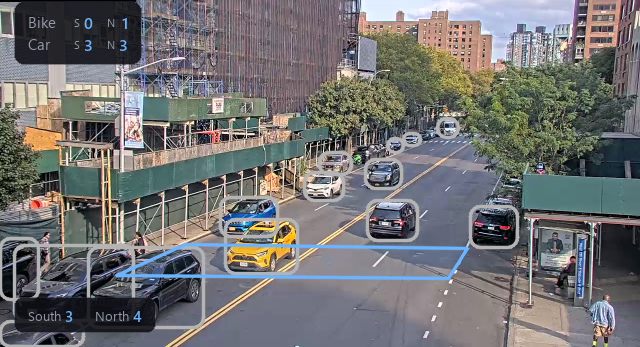}
    \end{minipage}

    \caption{Overview of the study network and traffic camera observations. 
    (a) Road network and camera locations; light blue lines indicate road segments monitored by traffic cameras. 
    (b) Traffic camera observation at location A. 
    (c) Traffic camera observation at location G.}
    \label{fig:combined_network_view}
\end{figure}

Large-scale urban road networks are characterized by closely spaced, signalized intersections with strong operational interdependence. Under high demand, these networks routinely exhibit queue formation, spillback, and downstream blocking, producing strongly coupled dynamics across adjacent road segments. Accurately characterizing the network-wide interactions therefore requires observations with sufficiently broad spatial coverage and temporal resolution. In dense environments such as Manhattan in New York City, traffic volumes are on the order of $10^6$ vehicle trips per day, yet network states are only partially observed due to the incomplete and spatially uneven coverage provided by GPS-based probe-vehicle trajectory data.  Crucially, the  penetration rate of such trajectory data are neither uniform nor stable. Commercial districts with intense taxi and ride-hailing activity can exhibit penetration rates above 15 $\%$, whereas residential areas dominated by private vehicles may fall below 5 $\%$. In Manhattan, the contrast between Midtown’s commercial core and the Upper East Side’s residential character exemplifies this spatial heterogeneity. Penetration rates also vary systematically over time as the traffic changes across hours, days, and weeks: morning and evening peaks are often dominated by commuters using navigation applications; midday traffic includes a larger share of taxis and delivery fleets with higher reporting rates; and weekends shift toward recreational travel with different participation in location-based services. As a result, the mapping between observed trajectories and the true underlying traffic volumes is time-varying and location-dependent, complicating any attempt to infer network-wide conditions from the trajectory data alone.
In this work, we focus on traffic volume estimation in the urban area of Manhattan, New York City, using two complementary data sources.  The primary source is the probe-vehicle information, provided by the company INRIX \cite{Inrix}, in the form of trajectories from a subset of vehicles traversing the road network. This subset represents only a fraction of the true traffic volume and therefore provides an incomplete, spatially biased view of the traffic conditions. To mitigate this partial observability, we incorporate measurements from nine traffic cameras that are sparsely distributed across the road network  (Fig. \ref{fig:combined_network_view}a).  To extract the traffic volume information,  we employed a computer vision pipeline that leverages New York City’s existing network of traffic cameras provided by NYC Traffic Management Center \cite{NYCTMC_Map}. The measurements were obtained using the object detection model (YOLO26 \cite{sapkota2025yolo26, yolo26_ultralytics}) optimized to process low-resolution (352 × 240 pixels) traffic camera footage (Fig. \ref{fig:combined_network_view},b \ref{fig:combined_network_view}c). 





\begin{figure}[t]
    \centering
    \begin{minipage}[t]{0.49\textwidth}
        \centering
        \vspace{0pt}
        \textbf{(a)}\\[0.3em]
        \includegraphics[width=\textwidth]{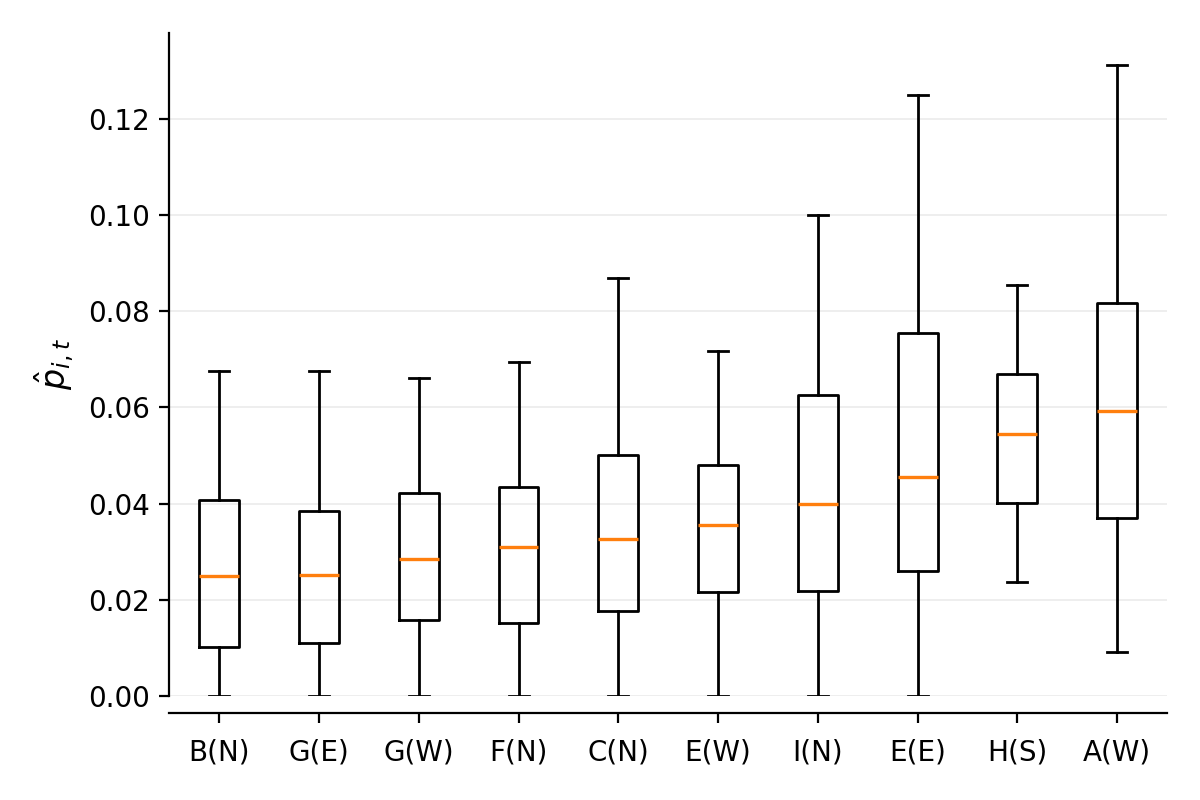}
    \end{minipage}
    \hfill
    \begin{minipage}[t]{0.49\textwidth}
        \centering
        \vspace{0pt}
        \textbf{(b)}\\[0.3em]
        \includegraphics[width=\textwidth]{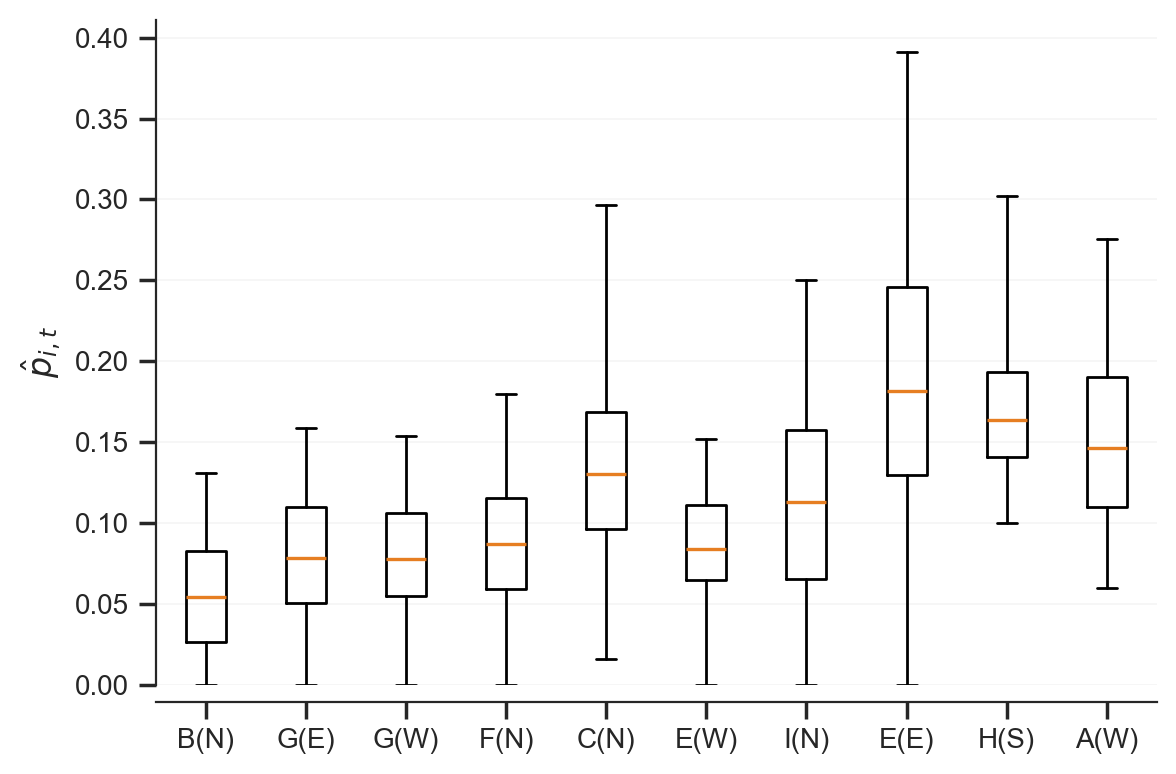}
    \end{minipage}
    \caption{Probe-vehicle penetration rates at traffic camera locations. The first letter indicates the camera location, and the letter in parentheses denotes the traffic direction, e.g., S = southbound. The boxes represent vehicle-weighted interquartile ranges (25th--75th percentiles) across 15-min time bins. (a) Penetration estimated using trajectory data from one month. (b) Penetration estimated after pooling trajectory data over three consecutive months within the same time-of-week bins.}
    \label{fig:penetration_rates}
\end{figure}

One of the key challenges in the primary source of information is the low penetration rate exhibited by the trajectory data. On average, probe trajectories account for less than  $5\%$ of the true traffic volume (Fig. \ref{fig:penetration_rates}a), while arterial road segments are represented by only a handful of measurements, making direct inference of road-level flows from trajectories alone highly unreliable. To deal with this challenge, instead of using real-time trajectory data, we analyze a history of the trajectories over several consecutive months to increase the trajectory sample size. Assuming approximately stationary traffic conditions within a given season, with no persistent network disruptions or structural shifts in demand, we aggregate probe-vehicle trajectories over three consecutive months. This pooling strategy increases the effective sample size and improves spatiotemporal coverage of  trajectories, thereby enabling more reliable network-wide traffic-state inference. In particular, such aggregation increases the effective penetration rate to as high as 20$\%$ (Fig. \ref{fig:penetration_rates}b). 
Using aggregated probe-vehicle trajectories, we formulate the traffic state estimation as a partially observed inference problem on an urban road network. Given the aggregated probe volumes and sparse traffic-camera observations, we seek a model that reconstructs the latent network traffic state and predicts segment-level traffic volumes at arbitrary time intervals. The objective is to estimate and forecast the full network state under severe, spatially nonuniform observability. Concretely, given a window of aggregated trajectory data and limited traffic camera observations, the objectives of the framework are to (i) infer the traffic state across all road segments in the network using sparse camera observations, (ii) estimate time- and location-dependent calibration factors that map trajectory-based traffic volume to camera observations, (iii) generate short-horizon forecasts of traffic volumes, and (iv) quantify predictive uncertainty over the inferred traffic volumes. 

\section{Methodology}
\begin{figure}
    \centering
      \includegraphics[scale=0.58]{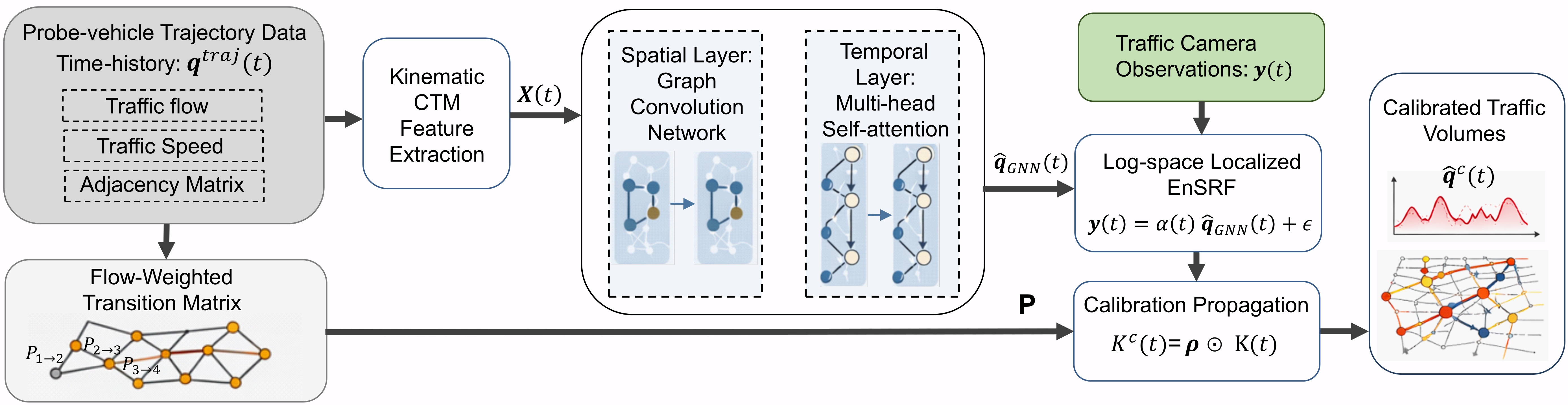}
      \caption{Overview of the proposed framework for network-wide traffic volume estimation. Probe-vehicle trajectories are converted into CTM-derived kinematic features and used as an input to a spatiotemporal GNN with graph convolution and temporal self-attention to obtain initial vehicle counts predictions. Sparse camera observations are then incorporated through a localized log-space EnSRF, while a flow-weighted transition matrix propagates calibration from observed to unobserved segments, producing the final calibrated traffic volume estimates.}
      \label{fig:Methodology}
\end{figure}

Traffic dynamics are considered to be governed by conservation laws and capacity constraints: vehicles move through a network of road segments, and the traffic density on each segment evolves according to the balance of inflow and outflow, while throughput is bounded by segment-specific limits determined by geometry and operating conditions. The CTM formalizes these principles within a discrete, network-based conservation framework, describing the temporal evolution of traffic states while capturing nonlinear congestion phenomena. Specifically, when downstream storage capacity is exhausted, upstream discharge is throttled, giving rise to queue formation and upstream-propagating congestion waves, a mechanism known as spillback.  Although the CTM was introduced in the context of freeway traffic, it is fundamentally a discrete-time, discrete-space conservation-law model with a Fundamental Diagram (FD) and explicit supply–demand constraints \cite{daganzo1994cell}. The model principles remain valid on urban links when traffic states are interpreted as mesoscopic, space–time averages over segments and time steps that are large enough to smooth signal-cycle microstructure \cite{daganzo1995cell}. 
However, despite the proven applicability of the CTM for modeling urban traffic,  real-world traffic dynamics exhibit substantial, context-dependent variability driven by recurrent cycles, exogenous perturbations, and complex network interactions that are difficult to parameterize explicitly. To capture these effects, we employ a GNN to learn data-driven traffic structure from probe-vehicle trajectories. Spatially,  the GNN performs message passing over the road network to model inter-segment dependencies and spillover effects. Temporally, it uses attention mechanisms to represent recurrent and nonstationary patterns by adaptively emphasizing the most informative past states for forecasting. 
A central challenge is that vehicle trajectories are an undersampled and biased view of true traffic conditions.  Penetration rates vary across space and time, so trajectory-derived counts are neither stable nor proportional estimators of true volume.  Under an unknown and potentially nonstationary sampling mechanism, fitting CTM flow parameters directly to probe counts would largely fit the measurement process (sampling bias) rather than the underlying traffic dynamics. We therefore avoid inferring absolute demand from undersampled trajectory counts and instead, extract CTM-consistent kinematic features from the aggregated trajectory data and calibrate quantities that are comparatively identifiable under FD assumptions, such as free-flow speed and congestion wave speed. In this context, the FD is treated as a closure relationship that maps speed to pseudo-density and pseudo-flow, from which we compute the CTM sending and receiving functions that define physically admissible traffic transitions. The GNN is then trained to learn structured deviations from this CTM baseline, capturing urban effects such as signal delay, turning friction, queue spillback, and intersection interactions, while remaining anchored to a conservation-consistent dynamics driven by speed measurements. By conditioning the learned component on CTM-derived kinematic features, the hybrid model acts as a physics-informed correction to CTM dynamics rather than an unconstrained surrogate. Model predictions are subsequently used as a dynamical prior in an EnSRF to calibrate traffic volumes using observations from a subset of cameras, after which the inferred calibration factors are propagated to road segments without camera coverage using a flow-weighted transition matrix derived from aggregated trajectory data.
Fig.\ref{fig:Methodology} illustrates a general overview of the proposed framework.

\subsection{Traffic Network Representation}
Consider an urban road network represented as a directed graph $G=(\mathcal{V},\mathcal{E})$,  where $\mathcal{V} = \{1, 2, \ldots, N\}$ is the set of road segments, and edges $\mathcal{E} \subseteq \mathcal{V} \times \mathcal{V}$ represent physical connectivity between segments. Each segment is characterized by length $l_i$, number of lanes $L_i$, and capacity $C_i$ (veh/hour), where  $i \in \{1, \ldots, N\}$ denotes segment index in the network. The network structure is further characterized through an adjacency matrix ${A}\in\mathbb{R}^{N\times N}$. Using the trajectory data from the probe vehicles, at each discrete time step $t$ (with interval $\Delta t$), we consider vehicle counts $q_i^{\text{traj}}(t)$ as a biased representation of the true traffic volume available for all road segments within a network. Our assumption is that while the counts $q_i^{\text{traj}}(t)$ are severely undersampled, speeds are accurately observed. We therefore reformulate CTM as a feature extractor that uses speed information to infer the traffic state. For each segment $i$ with observed speed $v_i(t)$ and estimated free-flow speed $v_i^{\text{free}}$, the normalized speed ratio indicates congestion level:
\begin{equation}
    b_i(t) = \frac{v_i(t)}{v_i^{\text{free}}} \in [0, 1],
\end{equation}
with $b_i(t)\approx 1$ corresponding to free-flow conditions and $b_i(t)\approx 0$ indicating severe congestion. Using the triangular FD, and establishing $v_{\text{crit}}$ as the speed at capacity, the speed is related to traffic density as:
\begin{equation}
\rho^d_i(t) = \begin{cases}
\dfrac{Q_i^{\max}}{v_i^{\text{free}}} \cdot \left(1 - b_i(t)\right) 
\cdot \dfrac{v_i^{\text{free}}}{v_i^{\text{free}} - v_w} 
& \text{if } b_i(t) \geq \dfrac{v_{\text{crit}}}{v_i^{\text{free}}} 
\\[10pt]
\rho_{\text{jam}} \cdot \left(1 - b_i(t)\right) .
\end{cases}
\end{equation}
Here $v_w$ is the wave speed of congestion propagation, and the threshold $v_{\text{crit}}/v_i^{\text{free}}$ is the normalized traffic flow speed at capacity \cite{newell1993simplified}; the $ Q^{max}_i=C_i \Delta t$ is the maximum traffic capacity of the road segment $i$ at each time interval $\Delta t$, and $\rho_{\text{jam}}$ is the jam density. 
The demand $D_i(t)$ and supply $S_i(t)$ functions are then estimated using the traffic density:
\begin{align}
    D_i(t) &= \min\left( \rho^d_i(t) \cdot v^{\text{free}}_i, \, Q^{max}_i \right) \\
    S_i(t) &= \min\left( v_w \cdot (\rho_{\text{jam}} - \rho^d_i(t)), \,  Q^{max}_i \right).
\end{align}
 The flow from segment $i$ to downstream segment $j$ is estimated as:
\begin{equation}
q_{i \to j} = \min\left(D_i \cdot \beta_{ij}, S_j \cdot \beta_{ij}\right),
\end{equation}
where $\beta_{ij}$ is the split/turn ratio from segment $i$ to $j$.  According to the CTM formulation,  the traffic flow on segment  $i$  at time $t+\Delta t$ is determined by (i) the traffic flow at the previous time step $q_i(t)$,  (ii) the inflow into the segment $ q_{j \to i}$,  (iii) the outflow from the segment $q_{i \to k}$, and (iv) the flow at the boundaries:
\begin{equation}
q_i^{\text{CTM}}(t+1) = q_i(t) + \sum_{j \in \mathcal{U}_i} q_{j \to i} - \sum_{k \in \mathcal{D}_i} q_{i \to k} + q_i^{\text{BC}}.
\end{equation}
The $\mathcal{U}_i$ and $\mathcal{D}_i$ denote upstream and downstream segment $i$ neighbors, and $q_i^{\text{BC}}$ accounts for normalized flows at the boundaries. To translate the CTM-based information into the GNN, we formulate a feature vector that encodes spatio-temporal traffic conditions and the boundary flows for each road segment $i$ as:
\begin{equation}
\mathrm{X}_i(t) = \left[q_i^{\text{traj}}(t), \mathbf{f}_i^{\text{temp}}(t), q_i^{\text{BC}}(t), \mathbf{f}_i^{\text{CTM}}(t)\right].
\end{equation}
The  $\mathbf{f}^{\text{temp}}$ encodes temporal features, including the hour of day $h$, the day of week $d$, and indicator variables, defined as: 
\begin{equation}
\begin{aligned}
\mathbf{f}^{\text{temp}}(t) = [ & \sin(2\pi h / 24), \cos(2\pi h / 24),\sin(2\pi d / 7), \\
                                &  \cos(2\pi d / 7), \mathbb{1}_{\text{weekend}}, \mathbb{1}_{\text{rush}}, \mathbb{1}_{\text{night}} ],
\end{aligned}
\end{equation}
where the first four terms encode smooth cyclical hour and day of the week, and the last three terms allow for a sharp traffic regime changes.   Unlike Fourier series commonly used in traffic prediction, this choice of temporal features can better capture the nonstationarity of real traffic scenarios in an urban network \cite{ye2022attention}.  The CTM kinematic features are encoded as $\mathbf{f}^{\text{CTM}} = [\vect{f}_i^{\text{sd}}(t); \vect{f}_i^{\text{sp}}(t)]$ with $\vect{f}^{\text{sd}}$ representing supply and demand conditions:
\begin{equation}
    \vect{f}^{\text{sd}}_{i}(t) = \begin{bmatrix}
        b_{i}(t) & \text{(speed ratio)} \\
        1 - b_{i}(t) & \text{(congestion level)} \\
        D_{i}(t) / C(i) & \text{(normalized demand)} \\
        S_{i}(t) / C(i) & \text{(normalized supply)} \\
        q_{i}(t) / C(i) & \text{(volume/capacity)} \\
        \text{LOS}_{i}(t) & \text{(level of service)} \\
        \mathbb{1}[b_{i}(t) < 0.5] & \text{(congested flag)} \\
        \mathbb{1}[0.7 < q_{i}(t)/C(i) < 0.9] & \text{(near-capacity flag)} \\
        q_{i}(t) / n_{\max}(i) & \text{(normalized count)}
    \end{bmatrix},
\end{equation}

and $\vect{f}^{\text{sp}}$ capturing the spatial context as:
\begin{equation}
    \vect{f}^{\text{sp}}_{i}(t) = \begin{bmatrix}
        \bar{b}^{\text{ds}}_{i}(t) & \text{(downstream speed ratio)} \\
        b_{i}(t) - \bar{b}^{\text{ds}}_{i}(t) & \text{(downstream gradient)} \\
        \bar{b}^{\text{us}}_{i}(t) & \text{(upstream speed ratio)} \\
        b_{i}(t) - \bar{b}^{\text{us}}_{i}(t) & \text{(upstream gradient)}
    \end{bmatrix}.
\end{equation}
The GNN learns spatio-temporal patterns from the features using two components: Graph Convolutional Network (GCN) to capture spatial dependencies in traffic propagation, and temporal self-attention layers to capture temporal dynamics.
Let $\mathrm{H}^{(0)}(\tau)\in\mathbb{R}^{N\times d}$ be the node embedding matrix at time $\tau$  for a model dimension $d$  after an input projection applied to  feature vector $\mathrm{X}_i(\tau)$. Spatial message passing is performed independently at each time step though $L_s$ graph-convolution layers. Denoting the node feature matrix on a spatial layer $\ell_s$ as $\mathrm{H}^{(\ell_s)}(\tau)\in\mathbb{R}^{N\times d}$,  each spatial layer updates the embeddings using the neighbor-aggregated message matrix $\mathrm{M}^{(\ell_s)}(\tau)$ and with respect to the adjacency matrix $\mathrm{A}$ as the following:
\begin{equation}
\label{eq:gnn_update}
\begin{aligned}
\mathrm{M}^{(\ell_s)}(\tau) &= \mathrm{A}\,\mathrm{H}^{(\ell_s)}(\tau),\\
\mathrm{H}^{(\ell_s+1)}(\tau) &= \sigma \Bigl( \mathrm{LN} \Bigl( 
\mathrm{M}^{(\ell_s)}(\tau)\mathrm{W}_{n}^{(\ell_s)} + \mathrm{H}^{(\ell_s)}(\tau)\mathrm{W}_{s}^{(\ell_s)} \\
&\quad + \mathrm{H}^{(\ell_s)}(\tau)\mathrm{W}_{r}^{(\ell_s)} \Bigr) \Bigr),
\end{aligned}
\end{equation}
where $\mathrm{W}_{n}^{(\ell_s)}$, $\mathrm{W}_{s}^{(\ell_s)}$, and $\mathrm{W}_{r}^{(\ell_s)}$ are learnable linear projections (neighbor, self, and residual projections, respectively), $\mathrm{LN}(\cdot)$ is LayerNorm, and  $\sigma$ is the Gaussian Error Linear Unit (GELU) activation function that introduces nonlinearity between layers.  After spatial processing, each segment $i$ has a sequence of latent embeddings over the history steps, denoted by $\mathrm{Z}_i \in \mathbb{R}^{H\times d}$ and constructed as:
\begin{equation}
\mathrm{Z}_i =
\begin{bmatrix}
\mathrm{H}^{(L_s)}(\tau_1)_{i,:}\\
\mathrm{H}^{(L_s)}(\tau_2)_{i,:}\\
\vdots\\
\mathrm{H}^{(L_s)}(\tau_H)_{i,:}
\end{bmatrix}.
\end{equation}
The transformer-style temporal blocks are applied to the embeddings:
\begin{align}
\tilde{\mathrm{Z}}_i &= \mathrm{Z}_i + \mathrm{MHA}\!\big(\mathrm{LN}(\mathrm{Z}_i)\big),\\
\mathrm{Z}_i &\leftarrow \tilde{\mathrm{Z}}_i + \mathrm{FFN}\!\big(\mathrm{LN}(\tilde{\mathrm{Z}}_i)\big),
\end{align}
where $\mathrm{MHA}(\cdot)$ is multi-head self-attention across time and $\mathrm{FFN}(\cdot)$ is a position-wise feedforward network.  
Here, each row of $\mathrm{Z}_i$ corresponds to one time step in the $H$-step history window and contains
a $d$-dimensional embedding of segment $i$ after spatial message passing. To obtain a road segment representation,  the temporal embedding matrix is flattened and then the learned projection is applied:
\begin{equation}
\vect{h}_i = \phi\!\left(\mathrm{W}_{\mathrm{out}}\,\mathrm{vec}(\mathrm{Z}_i) + \vect{b}_{\mathrm{out}}\right).
\end{equation}
Using two output heads, a $p$-step horizon of a mean increment and its uncertainty is then provided by the model:
\begin{align}
 \mu_{i,p} &= \big[\mathrm{MLP}_{\mu}(\vect{h}_i)\big]_p,\\
\sigma_{i,p} &= \exp\!\left(\big [\tfrac{1}{2}\mathrm{MLP}_{\log \sigma^2}(\vect{h}_i)\big]_p\right).
\end{align}

\subsection{Traffic State Estimation}
Using the GNN architecture described above, we train the model to learn network-level
structure from trajectories observations over  $t \in [0, T]$.  For learning and prediction,
we partitioned the dataset to form a history window of length $H$:
\begin{equation}
\vect{q}^{\mathrm{hist}}(t)=\big[q^{\mathrm{traj}}(\tau)\big]_{\tau=t-H+1}^{t}\in\mathbb{R}^{H\times N},
\end{equation}
and a future horizon of length $F$:
\begin{equation}
\vect{q}^{\mathrm{fut}}(t)=\big[q^{\mathrm{traj}}(t+p)\big]_{p=1}^{F}\in\mathbb{R}^{F\times N}.
\end{equation}
Under an assumption of heteroscedastic Gaussian distribution, the model predicts the future traffic state for the $p$-step increment in terms of the mean $\mu_{p}$ and variance $\sigma^2_{p}$:
\begin{align}
\Delta q_i(t+p) &\sim \mathcal{N}\!\big(\mu_{i,p},\ \sigma_{i,p}^2\big),
\qquad p=1,\dots,F.
\end{align}
Predicted counts are then reconstructed as:
\begin{equation}
\hat{q}^{\text{GNN}}_i(t+p)=q_i^{\mathrm{traj}}(t)+\mu_{i,p}.
\end{equation}
To learn the optimal parameters, the model is trained on a weighted sum of composite loss terms, given by: 
\begin{equation}
\mathcal{L}
= \lambda_{MAE} \mathcal{L}_{\mathrm{MAE}}
+ \lambda_{NLL}\mathcal{L}_{\mathrm{NLL}}
+\lambda_{cap}\mathcal{L}_{\mathrm{cap}}
+  \lambda_{cons}\mathcal{L}_{\mathrm{cons}}
.
\end{equation}
In the present methodology, the loss terms are combined with equal weights ($\lambda_{MAE}=\lambda_{NLL}=\lambda_{cap}=\lambda_{cons}=1$ ), treated as fixed hyperparameters, reflecting a deliberate balance between data fidelity and physical consistency.  This setting is consistent with multi-task learning formulations in which loss terms operate on comparable scales \cite{kendall2018multi}. 
The Mean Absolute Error (MAE) is computed as:
\begin{equation}
\mathcal{L}_{\mathrm{MAE}}
=\frac{1}{NF}\sum_{i,p=1}^{N,F}\left|q_i^{\mathrm{traj}}(t+p)-\hat{q}^{\text{GNN}}_i(t+p)\right|,
\end{equation}
and $\mathcal{L}_{\mathrm{NLL}} $ represents heteroscedastic Gaussian Negative Log-Likelihood:
\begin{equation}
\mathcal{L}_{\mathrm{NLL}}
=\frac{1}{NF}\sum_{i,p=1}^{N,F}\frac{1}{2}\left[
\log\!\big(\sigma_{i,p}^2\big)
+\frac{\big(q_i^{\mathrm{traj}}(t+p)-\hat{q}^{\text{GNN}}_i(t+p)\big)^2}{\sigma_{i,p}^2}
\right].
\end{equation}

The $\mathcal{L}_{\mathrm{MAE}}$ is used to fit the predicted mean traffic that is less sensitive to occasional spikes and counting noise. The $\mathcal{L}_{\mathrm{NLL}}$ term is included to learn segment specific heteroscedastic uncertainty in a likelihood-consistent way and to discourage trivial solutions that inflate variance, enabling uncertainty-aware calibration and confidence estimation.\\
The physics consistency is enforced through the capacity $\mathcal{L}_{\mathrm{cap}}$ and the conservation  $\mathcal{L}_{\mathrm{cons}}$ losses. When historical observations $\boldsymbol{q}^{\mathrm{hist}}$ are available, a global conservation of the network-wide vehicle total from the last observed step to the first forecast step is enforced through:
\begin{equation}
n^{\mathrm{tot}} = n^{\mathrm{hist}} + n_b,
\end{equation}
where,  $n^{\mathrm{hist}} = \sum_{i=1}^{N} q^{\mathrm{hist}}_i(t)$ is a total vehicles count on all the segments in the network at the last observed time, and $n_b=n_b^{\text{in}}-n_b^{\text{out}}$ is a net boundary flow, computed as the difference between vehicle counts entering and exiting the network at the boundaries. The conservation loss is therefore formulated as a violation of the traffic flow with the introduced tolerance band $\tau_b$:
\begin{equation}
\mathcal{L}_{\mathrm{cons}}
= 
\max\!\left(0,\left|\sum_{i=1}^{N} \hat{q}^{\text{GNN}}_i(t+1)-n^{\mathrm{tot}}\right|-\tau_b\right).
\end{equation}
It is important to note that the conservation is enforced as a global flux conservation for an entire road network, while the local flow interactions remain implicitly constrained by the adjacency matrix embedded in the GNN structure. The network-level conservation loss therefore complements rather than replaces local topological constraints, ensuring that the resulting  $\hat{q}^{\text{GNN}}$ remain globally traffic flow-consistent. 
This choice of modeling is based on the conditions for the CTM flux equalities, which would require accurately observed turning flows and boundary conditions that are unavailable under sparse and low-penetration probe-vehicle data. Enforcing conservation at the network level avoids over-constraining the model and forces $\mathcal{L}_{\mathrm{cons}}$  to act as a soft regularizer that stabilizes the prediction by penalizing implausible changes in the total vehicle count while still allowing local deviations consistent with the demand and supply functions.
Finally, for each segment in the road network,  the predicted vehicle counts are penalized if exceeding the capacity:
\begin{equation}
\mathcal{L}_{\mathrm{cap}}
=\frac{1}{NF}\sum_{i=1}^{N}\sum_{p=1}^{F}\max\!\left(0,\ \hat{q}^{\text{GNN}}_i(t+p)-Q_i^{\max}\right).
\end{equation}
In this work, the traffic state estimation is formulated as a spatio-temporal forecasting problem in which a CTM-informed GNN maps an $H$-step history of probe-trajectory counts to a probabilistic $F$-step prediction for each road segment. The model outputs a heteroscedastic Gaussian mean and uncertainty predicting count increments relative to the last observed step, and is trained end-to-end using data-fit losses together with soft regularizer that encourages physically plausible traffic states. 

\subsection{Traffic Calibration via Log-space Localized EnSRF}
To estimate a time-varying calibration field and update the model to be consistent with the traffic camera observations on the road network, the proposed framework employs a localized deterministic EnSRF under the assumption of a Gaussian modeling error. 
Let $y_{i,t}$ denote the high-fidelity camera observations available on a subset of road segments $\mathcal C_m\subset \mathcal{V}$.  
To estimate the calibration field, we first introduce a positive, time-varying calibration factor $\alpha_{i}(t)$ that maps the GNN predictions to the camera observations:
\begin{equation}
y_i(t) = \alpha_i(t)\,\hat q^{\text{GNN}}_i(t) + \epsilon_i, 
\qquad 
\epsilon_i\sim\mathcal{N}(0,\sigma_y^2),
\label{eq:obs_model_alpha}
\end{equation}
where $\hat q^{\text{GNN}}_i(t)$ is a trained GNN model prediction and $\epsilon_i$ represents camera measurement noise. To enforce positivity and obtain representative uncertainty, the calibration factor is estimated in log-space as $\beta_{i}(t)=\log\alpha_{i}(t)$.  We further factorize the calibration into interpretable temporal modulations:
\begin{equation}
\beta_{i}(t)=\beta^{\text{base}}_{i}+\beta^{\text{hour}}_{h(t)}+\beta^{\text{day}}_{d(t)}+
\beta^{\text{regime}}_{c_i(t)},
\end{equation}
where $h(t)\in\{0,\dots,23\}$ is hour-of-day, $d(t)\in\{0,\dots,6\}$ is day-of-week, and $c_i(t)$ is a coarse traffic regime index. In log-space, an approximately linear observation is formulated as the following:
\begin{equation}
\label{eq:log_ratio_obs}
\begin{aligned}
z_i(t) &:= \log(y_i(t)+\varepsilon) - \log(\hat q^{\text{GNN}}_i(t)+\varepsilon)\approx \beta_i(t) + \eta_i(t), \quad \eta_i(t) \sim \mathcal{N}(0, R_{z,i}(t)),
\end{aligned}
\end{equation}
where $\varepsilon>0$ is a constant introduced for numerical stability. The log-space observation variance is then approximated using a delta-method expression with a base noise floor $\sigma_0$:
\begin{equation}
R_{z,i}(t)=\sigma_0^2+\frac{\sigma_y^2}{(y_i(t)+\varepsilon)^2}.
\label{eq:rz}
\end{equation}
The log-calibration field is decomposed into the regime-specific components to capture systematic temporal variability and congestion-dependent bias. This structured parameterization promotes information sharing under sparse camera coverage and enables the calibration to generalize across future time steps, thereby maintaining multi-step forecasts on the correct volume scale while yielding more stable and interpretable uncertainty estimates. 
To predict the future traffic state, the filtering posterior of the calibration field is approximated by an ensemble $\{\beta^{(m)}(t)\}_{m=1}^\mathrm{M}$, where $\mathrm{M}$ is the number of members in the ensemble. For the brevity of notations, the forecast (prior) and analysis (posterior) ensembles are denoted by superscripts $f$ and $a$, respectively, where $\beta^{(m),f}(t)$ is equivalent to $\beta^{(m)}(t\mid t-1)$, and $\beta^{(m),a}(t-1)$ is to $\beta^{(m)}(t-1\mid t-1)$. 
To stabilize calibration on weakly observed arterial road segments, an Ornstein--Uhlenbeck (OU) forecast model is applied in log-space as:
\begin{equation}
\begin{aligned}
\beta^{\mathrm{base},(m),f}_i(t)
&= (1-\lambda_{\mathrm{base}})\,
   \tilde{\beta}^{\mathrm{base},(m),f}_i(t)
   + \lambda_{\mathrm{base}}\,\beta^\star(t)
   + \xi^{\mathrm{base},(m)}_i(t), \\
\beta^{r,(m),f}(t)
&= (1-\lambda_{\mathrm{glob}})\,
   \beta^{r,(m),a}(t-1)
   + \xi^{r,(m)}(t),
   \qquad r \in \{\mathrm{hour}, \mathrm{day}, \mathrm{regime}\}.
\end{aligned}
\label{eq:beta_forecast}
\end{equation}
where $\beta^\star(t)$ is the median of the baseline ensemble mean, ${Q}$ is the process-noise covariance, and  $\xi^{(\cdot)}(t)\sim\mathcal{N}({0},{Q}^{(\cdot)})$.  The $\lambda_{\mathrm{base}} \in (0,1)$ and $\lambda_{\mathrm{glob}} \in (0,1)$ are the mean-reversion rates controlling the speed at which the base calibration factor and the global temporal patterns revert toward their respective priors at each forecast step. Here, the $\lambda_{\mathrm{base}}<\lambda_{\mathrm{glob}} $, reflecting the asymmetry between the two components:  the base factor $\beta_i^{\mathrm{base}}$ is allowed to drift slowly to track gradual penetration rate changes, while the global temporal patterns $\beta^r$ are pulled more strongly toward neutral multipliers to prevent overfitting to sparse observations.\\
For a scalar observation on road segment $i$, the predicted log-ratio under an ensemble member $m$ is the effective log-calibration:
\begin{align}
\hat z_i^{(m),f}(t)
&= \beta^{\mathrm{base},(m),f}_i(t)
  + \beta^{\mathrm{hour},(m),f}_{h(t)}(t) \\
&\quad + \beta^{\mathrm{day},(m),f}_{d(t)}(t)
  + \beta^{\mathrm{tr},(m),f}_{c_i(t)}(t).
\end{align}
The predicted observation mean is then defined as  $\bar z_i^{f}$,  and used to estimate the corresponding  observation anomalies as  $\mathrm{Z}_i^{f}=[z_i^{(m),f}-\bar z_i^{f}]_{m=1}^{M}$.  Similarly,  the ensemble mean  $\bar\beta^{f}$  is used to estimate the forecast anomalies as $\mathrm{B}^{f}=[\beta^{(m),f}-\bar\beta^{f}]_{m=1}^{M}$.  These terms are then employed to form the ensemble-based variance $\mathrm{Var}(z_i)$ and covariance $\mathrm{Cov}(\beta,z_i)$. 
The Kalman gain is computed as:
\begin{align}
\mathrm{K}_i(t)
&= \frac{\mathrm{Cov}(\beta,\hat z_i)}
        {\mathrm{Var}(\hat z_i)+R_{z,i}(t)} \\
&= \frac{\mathrm{B}^{f}(t)\,\mathrm{Z}_i^{f}(t)^\top/(\mathrm{M}-1)}
        {\mathrm{Z}_i^{f}(t)\,\mathrm{Z}_i^{f}(t)^\top/(\mathrm{M}-1)+R_{z,i}(t)}.
\label{eq:ensrf_gain}
\end{align}
For each available observation $z_i(t)$, we form the innovation:
\begin{align}
\nu_i(t)=z_i(t)-\bar z_i^{f}(t).
\end{align}
The deterministic EnSRF update of the baseline mean and anomalies is then formulated as:
\begin{align}
\bar\beta^{\mathrm{base},a}(t) &= \bar\beta^{\mathrm{base},f}(t)+\mathrm{K}_i(t)\,\nu_i(t),\\
\mathrm{B}^{a}(t) &= \mathrm{B}^{f}(t)-\gamma_i(t)\,\mathrm{K}_i(t)\,\mathrm{Z}_i^{f}(t),
\label{eq:ensrf_update}
\end{align}
where $\gamma_i(t)$ is the standard square-root shrink factor, defined as:
\begin{equation}
\gamma_i(t)=\frac{1}{1+\sqrt{R_{z,i}(t)/(P_{zz}(t)+R_{z,i}(t))}},
\end{equation}
where $P_{zz}(t)=\mathrm{Var}(\hat z_i^{f}(t))$. The global components $\beta^{\mathrm{hour}},\beta^{\mathrm{day}}$, $\beta^{\mathrm{traffic}}$ are updated analogously, but with a reduced gain scale and a capped number of observations per time step to prevent unstable global corrections.
Uncertainty in the calibration is quantified from the ensemble predictions. For each segment $i$, we compute the ensemble mean and standard deviation in log-space, and map the calibration factor $\alpha_i^{(m)}(t)=\exp(\beta_i^{(m)}(t))$ to the physical space as:
\begin{align}
\bar{\alpha}_i(t) &= \frac{1}{M} \sum_{m=1}^{M} \exp\bigl(\beta_i^{(m)}(t)\bigr), \label{eq:mean_alpha} \\
\mathrm{Var}[\alpha_i(t)] &= \frac{1}{M-1} \sum_{m=1}^{M} \bigl(\exp(\beta_i^{(m)}(t)) - \bar{\alpha}_i(t)\bigr)^2. \label{eq:var_alpha}
\end{align}
\subsection{Flow-Weighted Transition Matrix and Calibration Propagation}
A central element of the proposed approach is the use of a flow-weighted transition matrix to propagate calibration factors from camera-observed segments throughout the road network. The transition matrix $\mathrm{P} \in \mathbb{R}^{N \times N}$ is defined as:
\begin{equation}
P_{ij} = P(\text{vehicle to } j \mid \text{vehicle is at } i) = \frac{q^{\text{traj}}_{(i \to j)}}{\sum_{k \in N^{+}_{(i)}} q^{\text{traj}}_{(i \to k)}},    
\end{equation}

where $ N^{+}_{(i)}$ denotes the set of downstream road neighbors of segment $i$. Calibration biases can correlate both along downstream flow (vehicles leaving \(i\) enter \(j\)) and along upstream influence (shared demand patterns). A bidirectional propagation kernel is therefore formed by symmetrizing the transition matrix and applying a propagation decay factor \(\gamma^{\text{PD}}\in(0,1)\):
\begin{equation}
\mathrm{W}
=\gamma^{\text{PD}}\cdot \frac{1}{2}\Big(\mathrm{P}+\mathrm{P}^\top\Big).
\end{equation}
To capture influence beyond immediate neighbors, multi-hop kernels are constructed iteratively as:
\begin{equation}
\mathrm{W}^{(2)} = \gamma^{\text{PD}}\,\mathrm{W}^2, \qquad \mathrm{W}^{(3)} = \gamma^{\text{PD}}\,\mathrm{W}^{(2)}\mathrm{W}.
\end{equation}
For each camera observed segment $i$, a weighted combination of hop estimates is computed using a flow-localization vector \(\boldsymbol{\rho}^{(i)}\in[0,1]^N\), constructed using a geometric hop decay influence:
\begin{equation}
\tilde{\boldsymbol{\rho}}^{(i)} = \mathrm{W}_{:,i} + \tfrac{1}{2}\,\mathrm{W}^{(2)}_{:,i} + \tfrac{1}{4}\,\mathrm{W}^{(3)}_{:,i}.
\end{equation}
During assimilation, the EnSRF gain vector is calibrated with the flow-localization as:
\begin{equation}
\mathrm{K}_i^c(t) =\boldsymbol{\rho}^{(i)}\odot \mathrm{K}_i(t),
\end{equation}
ensuring that a camera observation at segment $i$ updates calibration factors only at flow-reachable road segments, with influence decaying geometrically with network distance. Segments with no flow-reachable path within the hops receive a zero gain and are not directly updated during assimilation. Separately, at every forecast step, a spatial diffusion process continuously transfers calibration information across the network by nudging each segment's log-space calibration ensemble toward a weighted average of its flow-connected road segments:
\begin{equation}
\tilde{{\beta}}^{\text{base},(m),f}_{i}(t) = (1-s)\,{\beta}^{\text{base},(m),f}_{i}(t) + s\sum_{j} W^{\text{eff}}_{ij}\,{\beta}^{\text{base},(m),f}_{j}(t),
\end{equation}
where $s$ is the spatial smoothing coefficient and $\mathrm{W}^{\text{eff}}$  is the row-normalised propagation matrix with self-loops, constructed as:
\begin{equation}
W^{\text{eff}}_{ij} = \begin{cases}
W_{ij} & i \neq j \\
\max\!\left(0,\, 1 - \sum_{k \neq i} W_{ik}\right) & i = j.
\end{cases}
\end{equation}
Here $\mathrm{W}^{\text{eff}}$ is a random walk matrix that ensures the spatial diffusion is stable and flow-preserving. For road segments with flow connections, the diffusion redistributes calibration information from camera-observed segments, with the self-loop weight absorbing the remainder to normalise each row to one. For the flow-isolated segments, the self-loop weight equals one and the diffusion reduces to an identity operation, leaving their calibration state unchanged. Such segments are therefore updated neither through the localized EnSRF gain nor through spatial diffusion, and their calibration factors are governed entirely by the OU mean reversion toward the network-wide prior $\alpha^\star(t)=\operatorname{median}_{i=1,\dots,N} \bar{\alpha}_i(t)$.
This formulation provides a principled fallback: in the complete absence of flow connectivity and observational support, the model conservatively reverts isolated segments toward the network consensus rather than allowing unbounded drift.
A per-segment calibration confidence is derived from the EnSRF ensemble predictions. In this framework, we assume that confidence $\delta_i(t)$ in the traffic volume prediction is increasing as the ensemble coefficient of variation decreases:
\begin{equation}
\label{eq:Confidence}
\begin{aligned}
\mathrm{cv}_i(t) &= \frac{\sqrt{\mathrm{Var}[\alpha_i(t)]}}{\bar{\alpha}_i(t)}, \\
\delta_i(t) &= \max\left( \delta_i(t-1),\, \frac{1}{1 + \mathrm{cv}_i(t)} \right),
\end{aligned}
\end{equation}
with $\mathrm{\delta}_i(t)=1$ for camera-observed road segments. To ensure stable calibration on weakly observed road segments, the propagated segment-level estimate is regularized toward a network-wide prior $\alpha^\star(t)$ using a confidence-weighted shrinkage formulation:

\begin{equation}
\alpha^{\mathrm{c}}_i(t)
=
\mathrm{\delta}_i(t)\,\bar\alpha_i(t)
+
\big(1-\mathrm{\delta}_i(t)\big)\,\alpha^\star(t).
\label{eq:conf_blend}
\end{equation}
This formulation preserves the calibration where the posterior is well constrained, while reverting toward a conservative network-wide prior on weakly observed or distant segments, preventing unbounded drift in regions with limited observations. 
Finally, the calibrated vehicle count is then obtained as:
\begin{align}
\hat q^{\text{c}}_i(t)=\alpha^c_i(t)\,\hat q^{\text{GNN}}_i(t).
\end{align}
The reliability of the propagated calibration is quantified using a confidence state that combines ensemble uncertainty, flow-localized observational support, and temporal confidence decay. The ensemble coefficient of variation provides a local measure of calibration uncertainty, while the flow-localization weight limits the influence of camera observations to dynamically connected road segments. 

\section{Results}

\subsection{Dataset}
The performance of the proposed framework is evaluated using traffic data from Manhattan, New York City. The study network consists of 2,394 directed road segments, with lengths ranging from 50 m to 2,000 m. CTM-based features are derived from GPS-based trajectory data collected during September, October, and November 2025. The analysis is based on the INRIX probe-vehicle dataset, comprising 1,031,206 vehicle trips.
The trajectory data are aggregated into a representative one-month dataset and discretized into 15-minute intervals. Speed observations used to construct the kinematic feature vector are derived from the same aggregated dataset, ensuring temporal consistency between traffic volume and speed measurements. Prior to the model training, trajectories are map-matched to the road network to reduce the impact of missing observations and GPS measurement noise \cite{Patrick}. Ground-truth reference measurements are obtained from fixed traffic cameras at distinct locations in the network (Fig. \ref{fig:combined_network_view}a). The camera locations used for calibration were selected based on a network observability analysis, in which segments were ranked according to observability confidence derived from the flow-weighted propagation structure of the network. The observability analysis is described in Appendix A. 

\subsection{Traffic Volume Estimation}
\begin{figure}[t]
    \centering
    \includegraphics[width=0.59\textwidth]{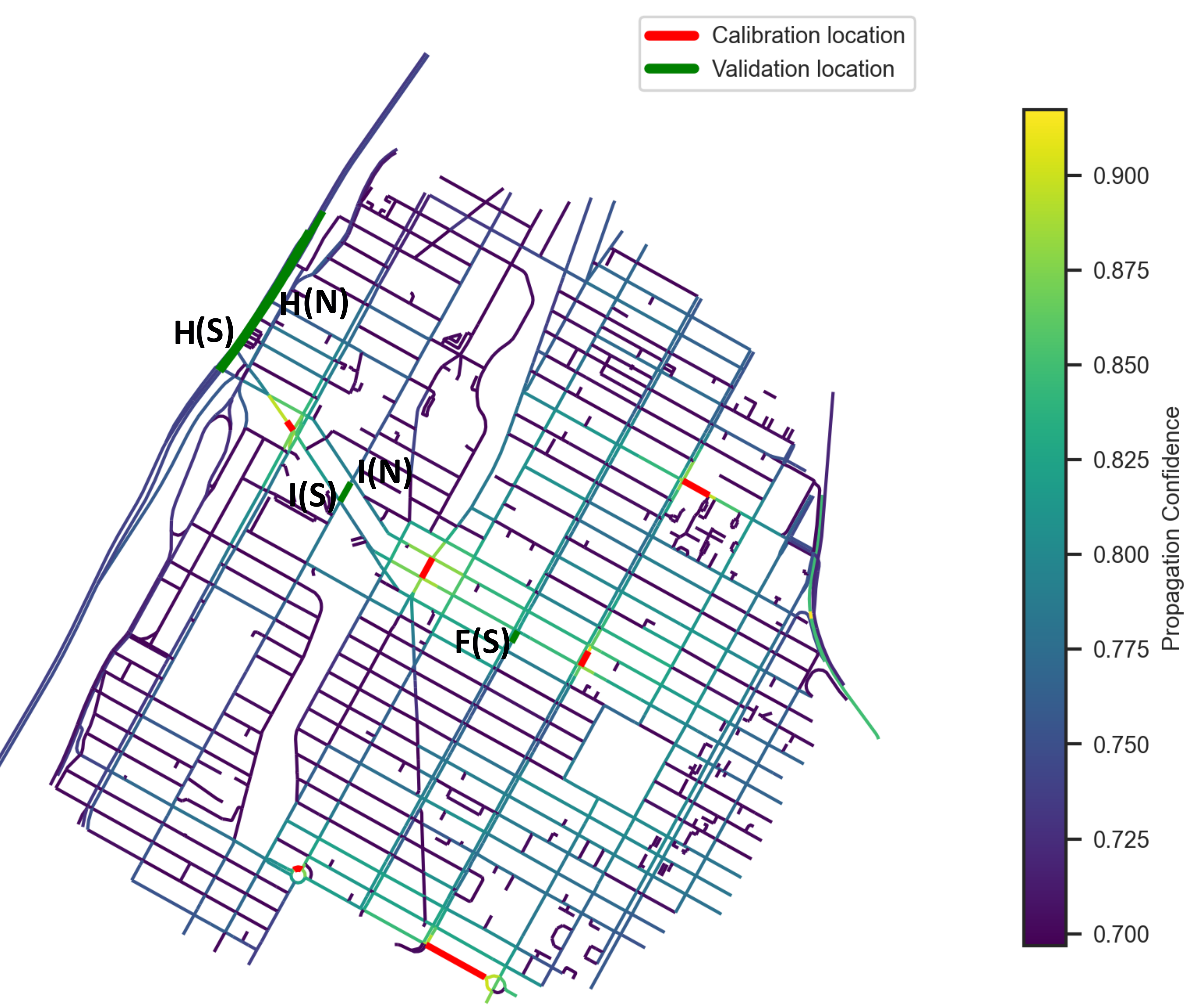}
    \caption{Spatial distribution of the propagated calibration confidence over the road network, with camera locations separated into calibration (red) and validation (green) subsets. Confidence reflects the strength of flow-connected influence and ensemble uncertainty after EnSRF assimilation. The first letter indicates the camera location, and the letter in parentheses denotes the traffic direction, e.g., S = southbound.}
    \label{fig:Res1}
\end{figure}

The primary objective is to estimate representative traffic volumes on the road network leveraging the aggregated trajectory data and sparse camera coverage. To assess both calibration performance and spatial generalization, a subset of five cameras is used to  estimate the time-varying calibration field, while the remaining four cameras are held out exclusively for out-of-sample validation. Fig.~\ref{fig:Res1} shows the camera-observed (red) and camera-unobserved (green) directed road segments, together with the corresponding propagation confidence estimated from the aggregated trajectories using Eq.~(\ref{eq:Confidence}). As expected, the confidence in the propagated calibration decreases with increasing distance from the cameras used for the calibration. This trend reflects the fact that calibration information is propagated mainly through the network along flow-connected segments, so its reliability gradually weakens as the topological and dynamical separation from directly observed camera locations increases.

\begin{figure}[t]
    \centering
    \includegraphics[width=0.59\textwidth]{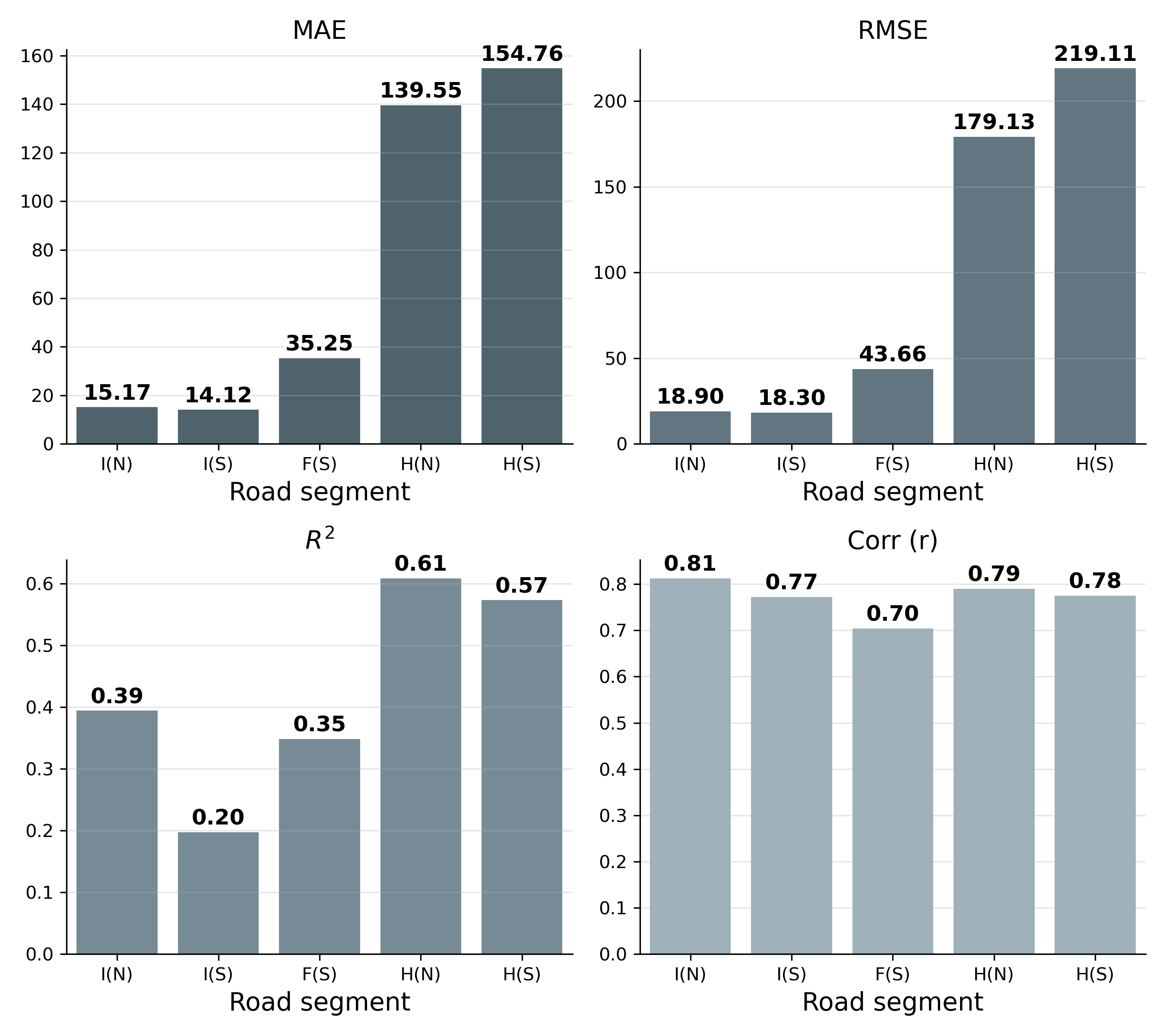}
    \caption{Traffic volume calibration accuracy at validation camera locations. The first letter indicates the camera location, and the letter in parentheses denotes traffic direction on a road segment in the network: S denotes southbound and N denotes northbound.}
    \label{fig:Res2}
\end{figure}
\begin{figure}[t]
    \centering
    \includegraphics[width=0.6\textwidth]{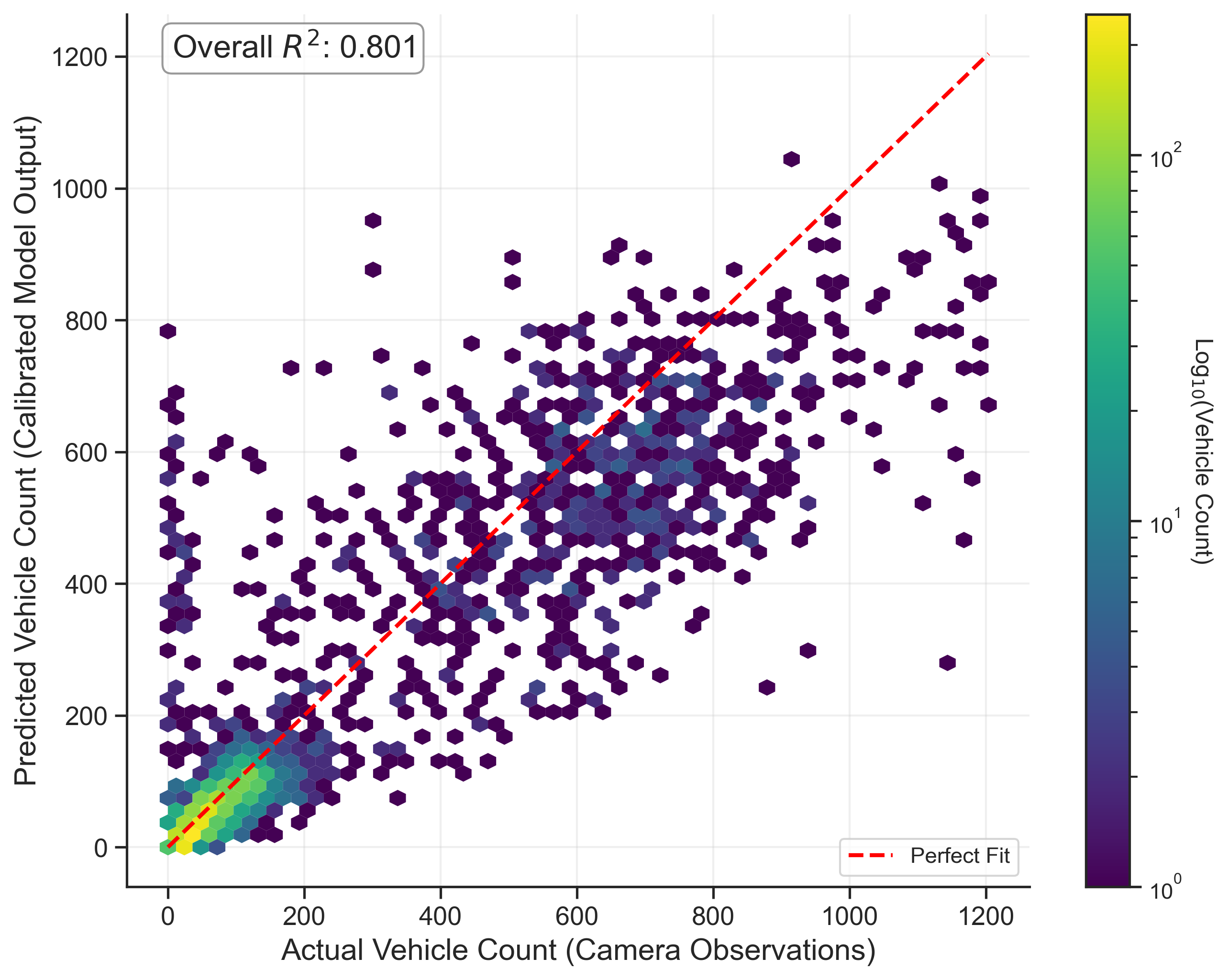}
    \caption{Model performance and error analysis across validation road segments. The hexbin plot shows the relationship between camera-observed traffic volumes and model-predicted vehicle counts across the validation set, with the overall \(R^2\) reported.}
    \label{fig:Res3}
\end{figure}
\begin{figure}[t]
    \centering
    \begin{minipage}[b]{0.47\textwidth}
        \centering
        \textbf{(a)}\\[0.3em]
        \includegraphics[width=\textwidth]{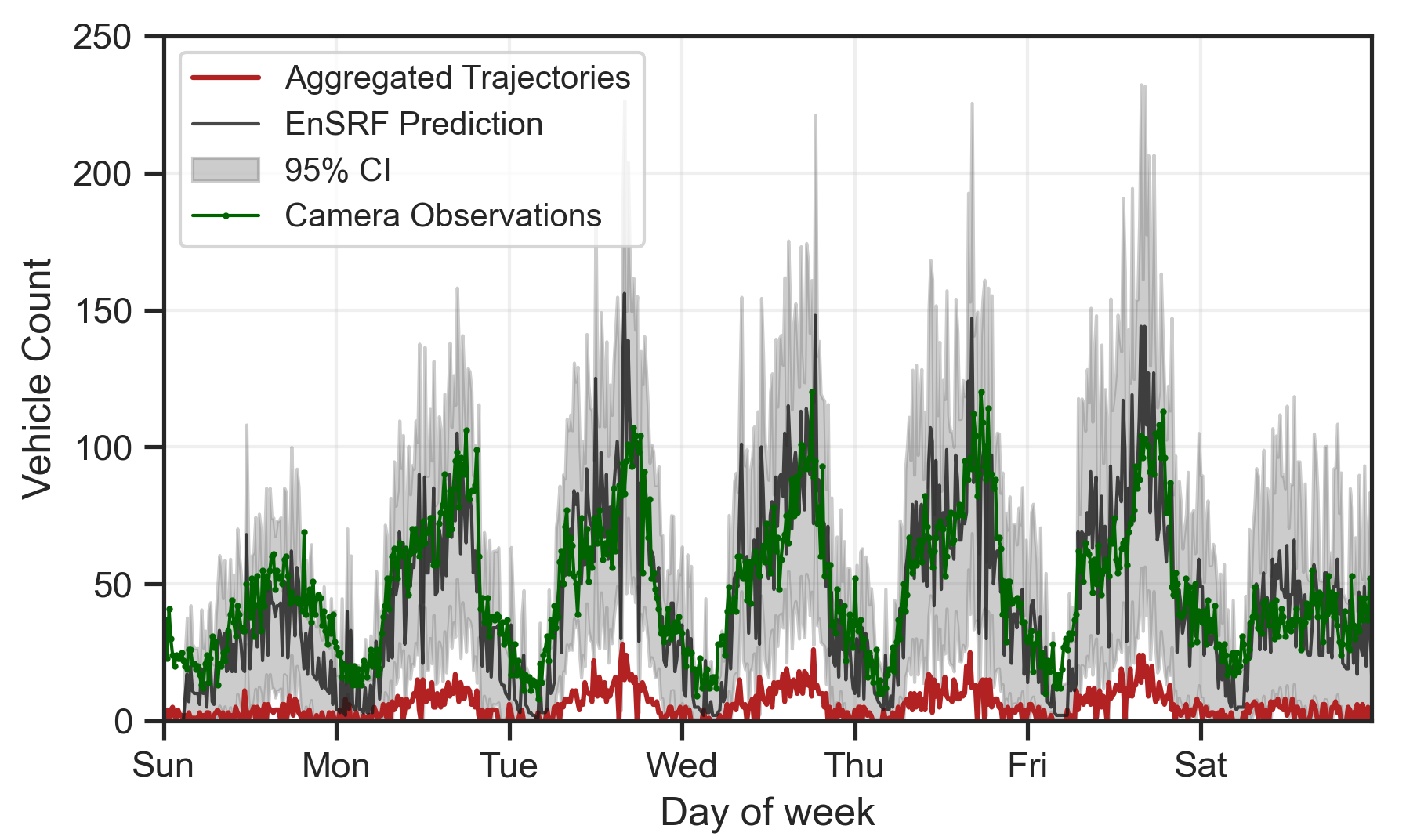}
    \end{minipage}
    \hfill
    \begin{minipage}[b]{0.47\textwidth}
        \centering
        \textbf{(b)}\\[0.3em]
        \includegraphics[width=\textwidth]{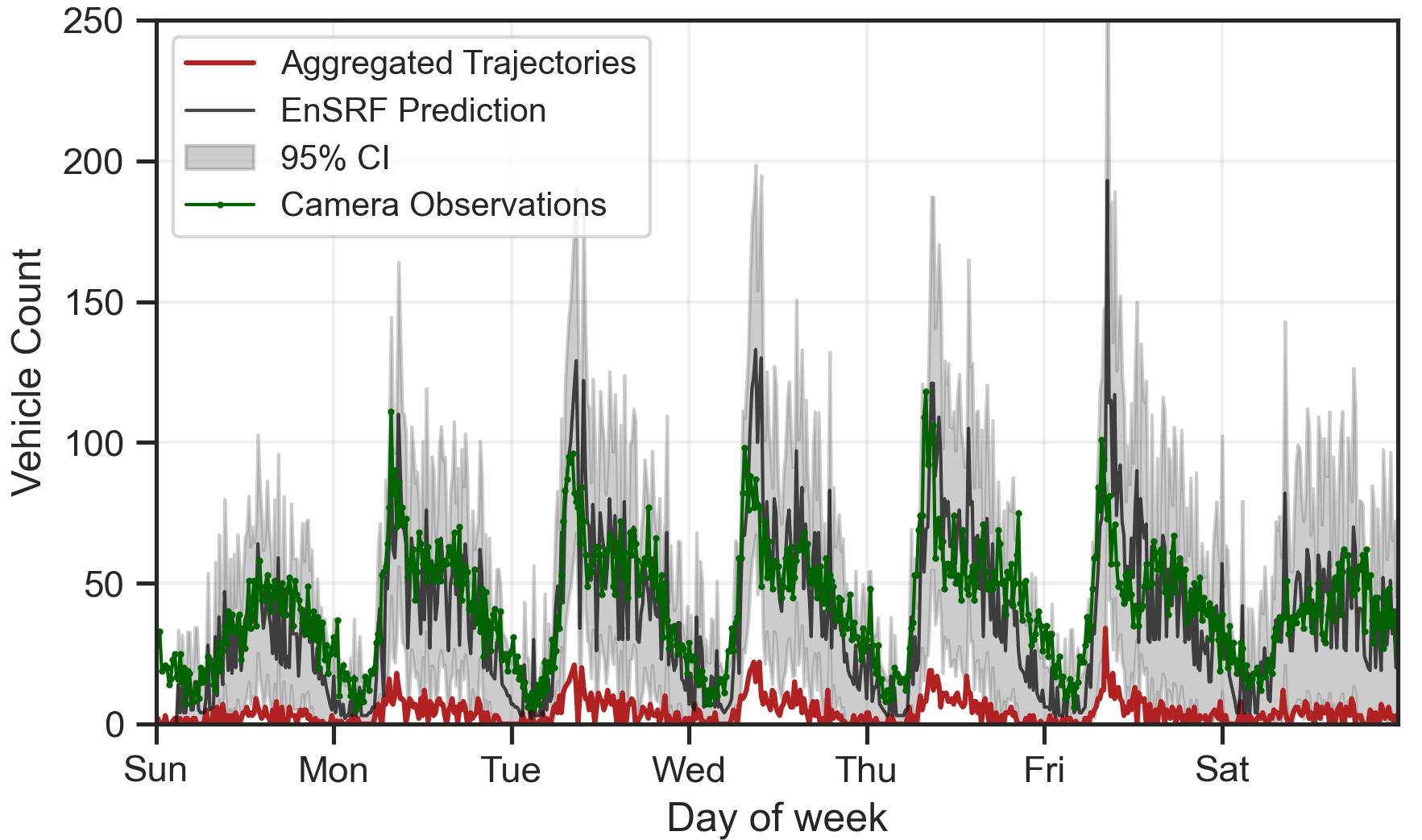}
    \end{minipage}
    \caption{Predicted traffic volumes at camera-unobserved locations. Results are shown for (a) northbound traffic at location I(N) and (b) southbound traffic at location I(S).}
    \label{fig:Res24}
\end{figure}
\begin{figure}[t]
    \centering
    \begin{minipage}[b]{0.49\textwidth}
        \centering
        \textbf{(a)}\\[0.3em]
        \includegraphics[width=\textwidth]{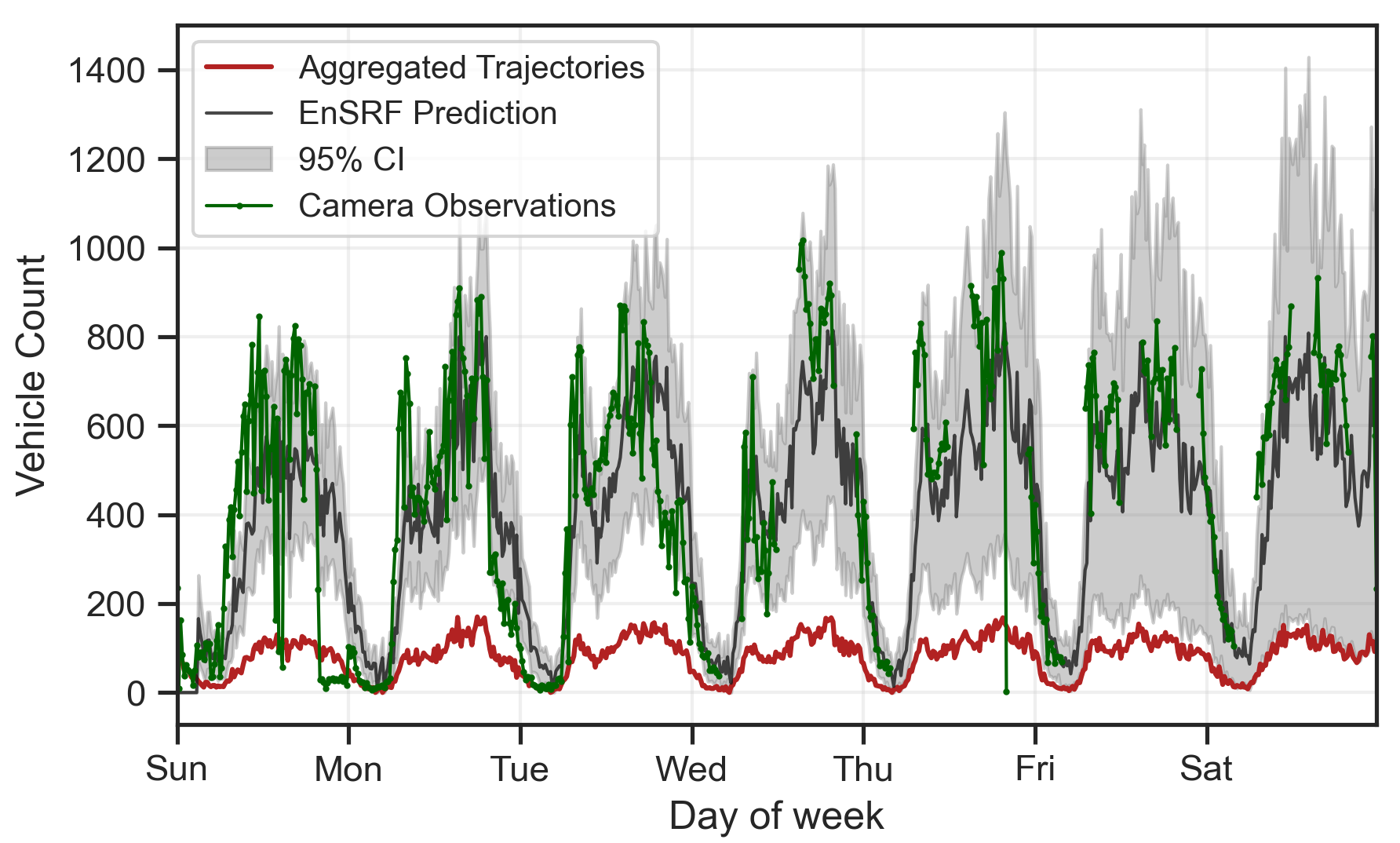}
    \end{minipage}
    \hfill
    \begin{minipage}[b]{0.49\textwidth}
        \centering
        \textbf{(b)}\\[0.3em]
        \includegraphics[width=\textwidth]{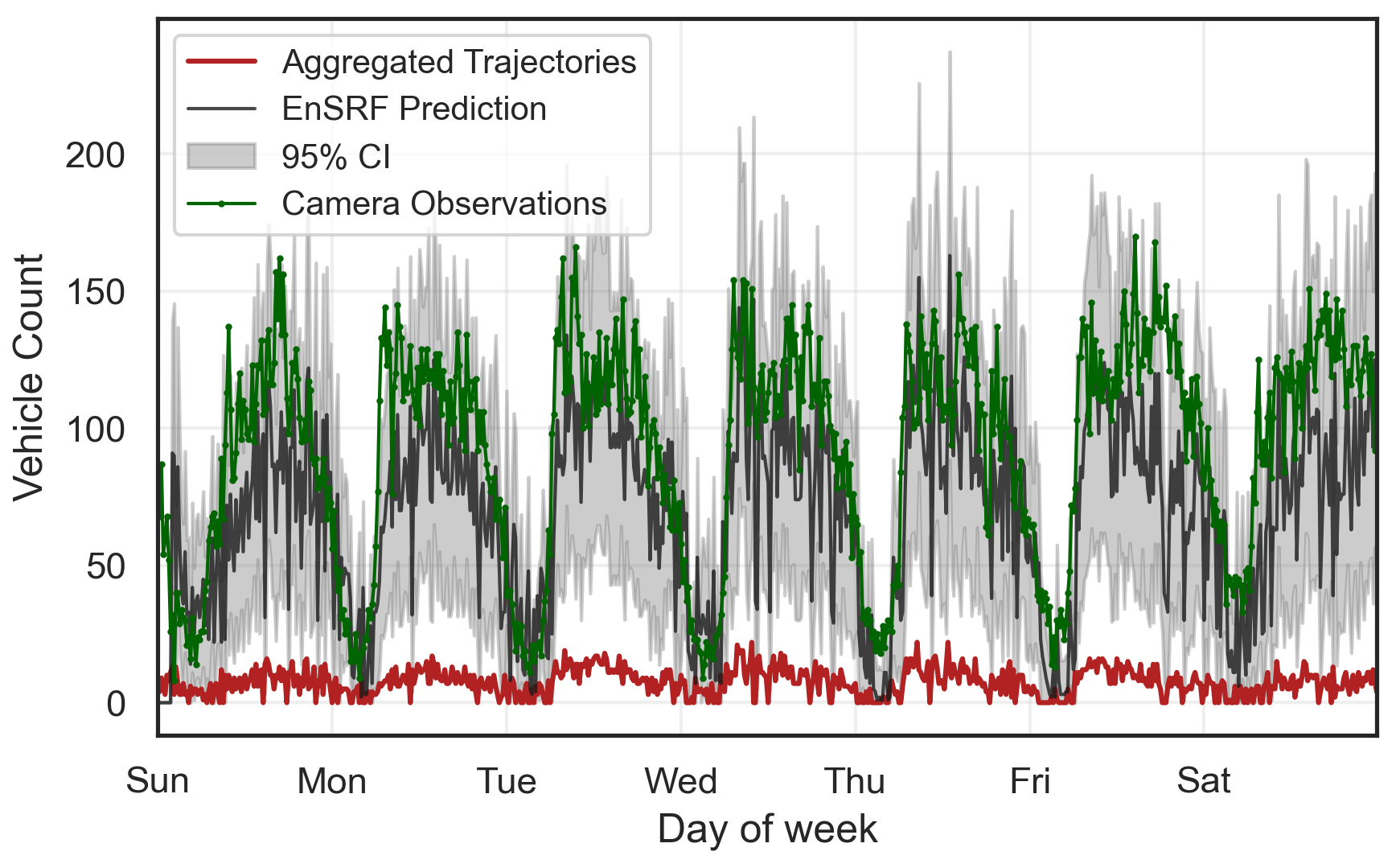}
    \end{minipage}
    \caption{Comparison of predicted traffic volumes at camera-unobserved arterial and freeway segments. Results are shown for (a) northbound traffic on freeway segment H(N) and (b) southbound traffic on arterial segment F(S). The intermittent drops in the camera observations are due to missing recording intervals and therefore reflect data gaps, not reductions in traffic volume.}
    \label{fig:Res22}
\end{figure}
\begin{figure}[t]
    \centering
    \includegraphics[width=0.99\textwidth]{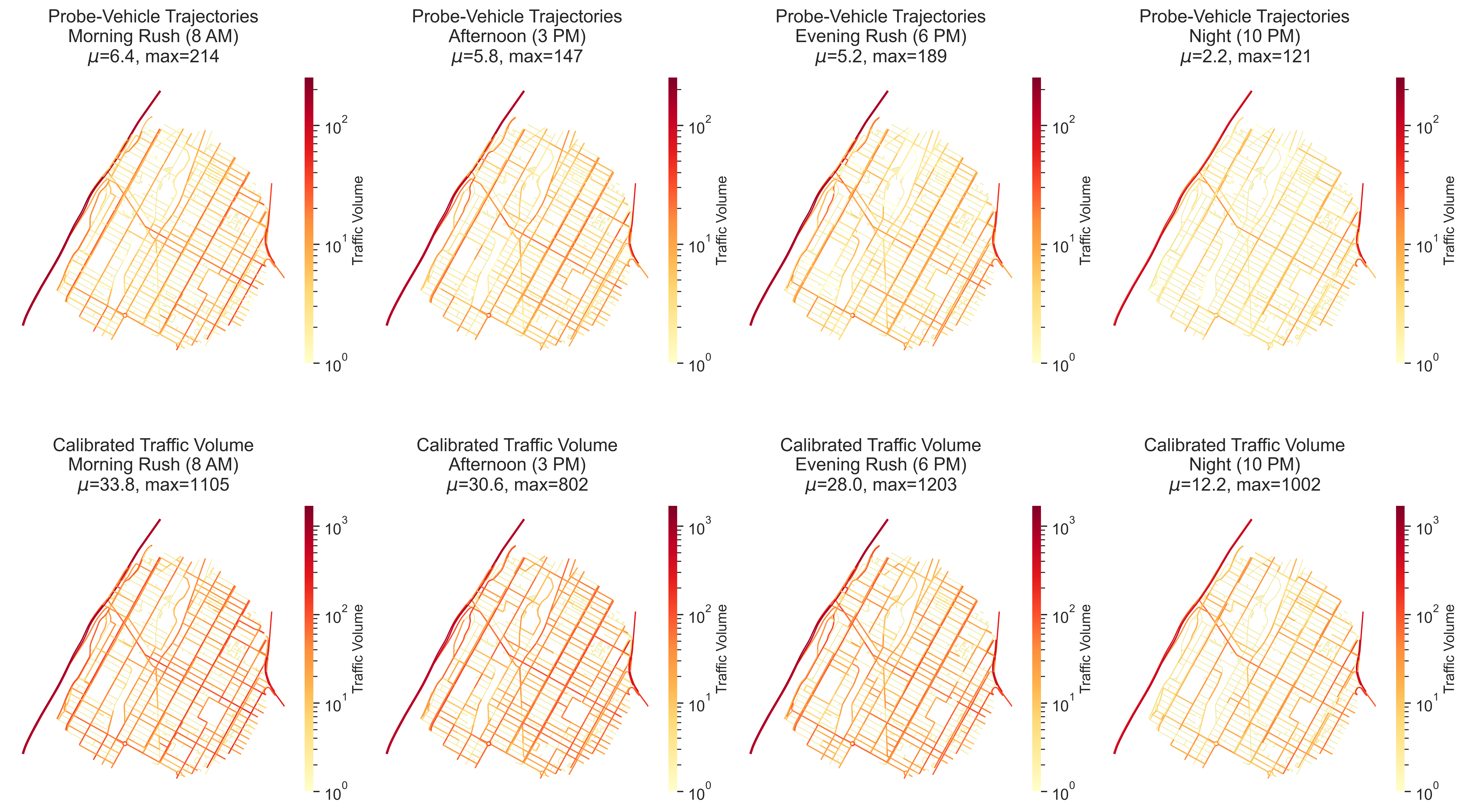}
    \caption{Comparison of probe-vehicle counts and calibrated traffic volume estimates across the road network at representative times of day: morning rush (8 AM), afternoon (3 PM), evening rush (6 PM), and night (10 PM).}
    \label{fig:Res4}
\end{figure}

The accuracy of the calibration is quantified using mean absolute error (MAE), root mean squared error (RMSE), the coefficient of determination ($R^2$), and the Pearson correlation coefficient ($r$).  The metrics are computed for each held-out camera locations.
As shown in Fig.~\ref{fig:Res2}, the calibrated predictions exhibit positive $R^2$ across all
camera-unobserved segments, indicating predictive superiority relative to a mean-baseline model. For arterial road segments I(N), I(S), and F(S), the model yields relatively low MAE and RMSE and strong correlation with the ground truth traffic volumes, suggesting that it captures the dominant variations in traffic dynamics, including both overall magnitude and temporal evolution.  This is further supported by the one-week traffic-volume forecasts shown in Fig. \ref{fig:Res24}.
The substantially larger errors on the freeway segments H(N) and H(S) are in part a direct consequence of the higher traffic volumes on these facilities, which are on the order of 900 vehicles (Fig. \ref{fig:Res22}a) compared to approximately 150 vehicles on an arterial segment (Fig. \ref{fig:Res22}b).  Since the MAE and RMSE are absolute metrics, larger volume ranges mechanically yield larger absolute errors even under comparable relative fit. This scale dependence is corroborated by the positive $R^2$ and sustained temporal correlation on the freeway segments, which are scale-invariant measures and indicate that the model preserves the relative temporal structure of traffic fluctuations despite the larger absolute residuals. The residual magnitude bias beyond the volume effect is attributable to weaker calibration constraints on these corridors, as freeway segments have no direct flow connectivity to the camera-observed arterial locations and receive only indirect updates through the network-wide calibration prior.
Fig.  \ref{fig:Res3} summarizes the calibration performance ($R^2=0.801$) averaged over all camera observations, indicating that predicted volumes scale consistently with camera observed traffic volumes across the full dynamic range. The higher point density at low-to-moderate counts reflects the predominance of urban arterial segments under typical operating conditions, where agreement between ground truth and model predictions is strongest. Dispersion increases at high counts, concentrated during peak-demand periods when traffic exhibits greater variability and the combined effect of measurement noise and model mismatch is amplified.  \\
Importantly, the time-series shown in Fig.~\ref{fig:Res24} and Fig.~\ref{fig:Res22} highlight the severe undersampling associated with low probe-vehicle penetration rates, for which even aggregated trajectory data represent only a small fraction of the true traffic volumes. Nevertheless, the proposed framework is able to align the model predictions into closer agreement with camera observations, including at camera-unobserved locations, thereby demonstrating its capability to recover physically plausible traffic volumes from sparse, low-penetration probe data. The 95\% prediction intervals shown in Fig.~\ref{fig:Res24} and Fig.~\ref{fig:Res22} reflect the segment-level uncertainty quantified by the EnSRF ensemble and reveal an important asymmetry between facility types. On the arterial road  segments, the prediction intervals are relatively uniform and the camera observations fall predominantly within the confidence band, indicating well-calibrated uncertainty at locations with strong flow connectivity to camera-observed segments. In contrast, on a freeway segment H(N) (Fig. \ref{fig:Res22}a), the prediction intervals widen progressively toward the end of the forecasted time-window. This behavior reflects the confidence decay mechanism embedded in the EnSRF forecast step: in the absence of direct camera observations or strong flow-connected calibration updates, segment-level confidence decays geometrically at each time interval, causing the ensemble spread to grow as the distance from the last informative update increases. However, despite this growing uncertainty, the predicted mean remains on a scale consistent with camera observations during peak hours, suggesting that the calibration factor itself remains reasonable even as its precision degrades. 
Finally, Fig. \ref{fig:Res4} illustrates the network-wide effect of sparse camera-based calibration on traffic volume estimation across four representative regimes: morning rush, afternoon, evening peak, and night. The aggregated trajectory-based counts are systematically underestimated across all four time windows, with the magnitude of bias varying by time-of-day, reflecting the non-stationary nature of probe-vehicle penetration rates. Following assimilation of the calibration factors, traffic volumes are elevated to a scale of camera observations, as reflected by the increase in network-wide mean and maximum values. The calibrated traffic volumes remain spatially coherent across corridors beyond the camera locations, demonstrating that the flow-localized gain propagates calibration factors to unobserved road segments in a manner consistent with network topology and traffic demand. 

\section{Conclusion} 
This study presents a framework for uncertainty-aware traffic volume estimation using complimentary information from probe-vehicles and traffic cameras. The proposed model integrates a Graph Neural Network (GNN) trained on probe-vehicle trajectory data with a Cell Transmission Model (CTM) serving as a physics-based feature extractor that provides congestion regime, supply-demand conditions, and wave propagation characteristics as inputs to the GNN.  Traffic volume estimation is performed using a localized Ensemble Square-Root Kalman Filter (EnSRF), which estimates segment-specific, time-varying calibration factors to account for heterogeneous and nonstationary probe-vehicle penetration rates, a practical challenge in traffic monitoring. The ensemble-based formulation further enables uncertainty quantification, allowing prediction intervals to be constructed for probabilistic assessment of traffic conditions across the road network.  \\
An important contribution of this work is the introduction of a flow-weighted transition matrix  that governs how calibration information is propagated from camera-observed road segments to the remainder of the network within the EnSRF update. This topology-informed transition matrix transfers appropriate calibration factors to camera-unobserved road segments in a manner consistent with network connectivity and dominant traffic patterns, improving generalization beyond the cameras field of view. Combined with confidence-weighted shrinkage toward a network-wide prior, the framework ensures that calibration estimates at unobserved segments remain physically plausible rather than diverging in the absence of direct observations. \\
The present study focuses on traffic volume estimation for the typical-week traffic operating conditions. Future work will extend the framework to explicitly account for exogenous disruptions, including public holidays, severe weather, incidents, and construction, through additional context inputs and event-conditioned dynamics. Other directions include refining intersection and signal effects in the dynamical prior, evaluating transferability across multiple urban networks, and moving toward scalable, continuously updated digital-twin representations of network traffic states. Although the methodology presented here leverages aggregated probe-vehicle data to improve the coverage and stability of the learned dynamics, the same framework is directly applicable to online traffic-volume estimation, provided that streaming trajectory data offer sufficient spatiotemporal coverage to support reliable state estimation and calibration updates. 

\section*{Acknowledgments}
This work was supported by the Center for Smart  Streetscapes, 
an NSF Engineering Research Center, under grant agreement EEC-2133516.\\
The authors gratefully acknowledge the NYC Traffic Management Center for providing access to traffic camera footage, with special thanks to Mohamad Talas, Rachid Roumila, Evgeniy Kudinov, and Mohammed Belal Uddin.  The traffic data were provided by INRIX company as part of the 2025 INRIX x MetroLab Challenge.

\bibliography{references}

\section*{Appendix A. Observability Analysis}
\label{app:observability}
\subsection*{A.1 Theory} 
Consider a road network characterized by $N$ directed segments. Let
$\bm{x}(t)\in\mathbb{R}^N$ denote the traffic state at discrete time $t$, where $x_i(t)$ is the number of vehicles on a road segment $i$. For observability analysis, we consider a Linear Time-Invariant (LTI) approximation obtained by linearizing the Cell Transmission Model (CTM) around a nominal trajectory under a fixed regime and  turning ratios:
\begin{equation}
\begin{aligned}
\bm{x}(t+1) &= \bm{A}\bm{x}(t) + \bm{B}\bm{u}(t), \\
\bm{y}(t)  &= \bm{C}\bm{x}(t),
\end{aligned}
\label{eq:state_obs_app}
\end{equation}
where $\bm{y}(t)\in\mathbb{R}^m$ are traffic state observations from $m$ traffic cameras,  $\bm{u}(t)\in\mathbb{R}^p$ denotes known exogenous inputs (boundary demands and on-ramp traffic flows), and  $\mathbf{B}\in\mathbb{R}^{N\times p}$ is  the input distribution matrix that maps the input to the segments where vehicles physically enter or leave the network.  The matrix $\bm{A}\in\mathbb{R}^{N \times N}$ is the effective linear propagation operator, encoding how perturbations in upstream traffic states influence downstream states over one time step under the assumed regime. Since the CTM influence graph differs between free-flow (information advects forward) and congestion (spillback propagates upstream), $\bm{A}$ is regime-dependent.  Assuming traffic cameras observe a set of segments $\{s_1,\dots,s_m\}$, and each camera provides a measurement of the vehicle density on its co-located road segment. The observation matrix $\bm{C}\in\mathbb{R}^{m\times N}$ is then a selection matrix:
\begin{equation}
    \bm{C} =
    \begin{bmatrix}
        \bm{e}_{s_1}^\top\\
        \vdots\\
        \bm{e}_{s_m}^\top
    \end{bmatrix},
\end{equation}
with $\bm{e}_j$ the $j$-th standard basis vector of $\mathbb{R}^N$.
For an LTI system, the pair $(\bm{A},\bm{C})$ is observable if the initial state $\bm{x}(0)$ can be uniquely recovered from a finite output sequence $\{\bm{y}(t)\}$ given known inputs $\{\bm{u}(t)\}$. Observability is then equivalent
to the rank condition $\operatorname{rank}(\mathcal{O})=N$, where the observability matrix is formulated as:
\begin{equation}
    \mathcal{O} \;\triangleq\;
    \begin{bmatrix}
        \bm{C}\\
        \bm{C}\bm{A}\\
        \bm{C}\bm{A}^2\\
        \vdots\\
        \bm{C}\bm{A}^{N-1}
    \end{bmatrix}
    \in \mathbb{R}^{mN\times N}.
    \label{eq:obs_matrix_app}
\end{equation}
In urban traffic networks, full observability is rarely achievable with sparse fixed cameras. We therefore estimate a rank-based observability index defined as:
\begin{equation}
    \gamma_{\mathrm{rank}} \;\triangleq\;
    \frac{\operatorname{rank}(\mathcal{O})}{N}\in[m/N,\,1],
    \label{eq:coverage_ratio_app}
\end{equation}
This quantity summarizes how many independent state directions are theoretically recoverable from the camera observations under the assumed linear dynamics. A complementary and numerically convenient metric is provided by the finite-horizon observability Gramian:
\begin{equation}
    \bm{W}_o(T) \;=\; \sum_{t=0}^{T-1} (\bm{A}^\top)^t\,\bm{C}^\top\bm{C}\,\bm{A}^t,
    \label{eq:gramian_finite_app}
\end{equation}
which satisfies $\bm{W}_o(T)=\mathcal{O}_T^\top\mathcal{O}_T$, where $\mathcal{O}_T$ is the observability matrix truncated to its first $T$ block
rows ($\bm{C},\bm{C}\bm{A},\dots,\bm{C}\bm{A}^{T-1}$).  When $\bm{A}$ is Schur-stable ($\rho(\bm{A})<1$) \cite{hespanha2018linea, horn2012matrix}, the infinite-horizon Gramian:
\begin{equation}
    \bm{W}_o \;=\; \sum_{k=0}^{\infty}(\bm{A}^\top)^k\,\bm{C}^\top\bm{C}\,\bm{A}^k
\end{equation}
exists and is the unique solution of the discrete Lyapunov equation:
\begin{equation}
    \bm{W}_o = \bm{A}^\top \bm{W}_o \bm{A} + \bm{C}^\top\bm{C}.
\end{equation}
If, in addition, $(\bm{A},\bm{C})$ is observable, then $\bm{W}_o\succ 0$. The
Schur-stability assumption is reasonable for the linearized CTM: in free-flow,
density perturbations propagate forward and dissipate at downstream
boundaries; in congestion, they propagate backward and dissipate at upstream
boundaries. Since the graph network constructed based on the physical description of the actual road network, the diagonal entries of $\bm{W}_o(T)$ provide a meaningful per-segment observability score:
\begin{equation}
    \mathrm{obs}_i \;\triangleq\; [\bm{W}_o(T)]_{ii}, \qquad
    \mathrm{conf}^{\mathrm{Gram}}_i \;\triangleq\;
        \frac{\mathrm{obs}_i}{\max_j \mathrm{obs}_j}\in[0,1].
    \label{eq:conf_from_gramian_app}
\end{equation}
Thereby, road segments with $\mathrm{obs}_i\approx 0$ are effectively unobserved by the camera placement over the horizon $T$ under the linear assumed dynamics. 

\subsection*{A.2 Justification of the Observability Surrogate}
\label{app:observability_justification}
The classical rank test and Gramian-based metrics are defined for linear
state--space systems. The CTM is generally nonlinear (with regime
switches between free-flow and congestion) and may be time-varying due to
fluctuating boundary demands and capacity constraints. We therefore interpret
\eqref{eq:state_obs_app} as a local surrogate obtained by linearizing the CTM
update around a nominal trajectory.  This approximation suffices for the objective: not to claim strict global
observability of the full nonlinear CTM, but to understand which camera locations are the most informative for the traffic state estimation. We emphasize that what is ultimately required for model parameter
estimation, if states cannot be reconstructed, parameters governing those states cannot be reliably identified from the available cameras measurements.
The observability matrix $\mathcal{O}$ and the Gramian $\bm{W}_o(T)$ quantify how state perturbations are transmitted to model outputs through repeated application of the dynamics. For traffic networks, this aligns with the
physics of conservation and advection: information about densities and flows propagates along directed connectivity, and its influence attenuates as it traverses intermediate segments and merges.
As the $\bm{A}$ differs in free-flow and congestion, we compute Gramian for the dominant regime observed during the calibration window for each road segment. When the nominal trajectory traverses both regimes,
the per-segment score is taken as the maximum across the regime-specific Gramian; this is conservative in the sense that a segment is counted as observable if it is observable in either regime.
When the dynamics vary continuously within a regime, the same idea extends with a time-varying linearization $\bm{A}(t)$. Defining the state-transition product
$\Phi(t,0) = \bm{A}(t-1)\bm{A}(t-2)\cdots \bm{A}(0)$ (so that $\bm{x}(t)=\Phi(t,0)\bm{x}(0)$ in the linearization), the finite-horizon 
time-varying observability Gramian is:
\begin{equation}
\bm{W}_o(T) \;=\; \sum_{t=0}^{T-1} \Phi(t,0)^\top \bm{C}^\top \bm{C}\,\Phi(t,0),
\end{equation}
which quantifies how strongly each state direction is expressed in the
measurement sequence over horizon $T$.  In practice, we use the Gramian-derived segment score
$\mathrm{conf}^{\mathrm{Gram}}_i$ as the primary calibration coverage metric:  road segments with very small $[\bm{W}_o(T)]_{ii}$ are weakly constrained by available traffic cameras and are expected to exhibit higher posterior uncertainty
after calibration. 

\subsection*{A.3 Case Study}
We apply the observability analysis to the Upper West Side corridor in New York City, spanning approximately 100th St to 145th St.  The network comprises \(N = 2{,}394\) directed segments, with only $ 13$ road segments directly observed by traffic cameras ($m/N \approx 0.54\%$). Given this sparse sensor deployment, full observability is not expected. However, the analysis reveals substantial partial observability: a large fraction of segments attain nonzero confidence, with the highest confidence concentrated at road segments near camera locations and along dominant traffic propagation paths as shown in Fig. \ref{fig:Res1}.

\end{document}